\documentclass[10pt]{wlscirep}
\usepackage[utf8]{inputenc}
\usepackage[T1]{fontenc}
\usepackage{lineno}
\usepackage{xcolor}
\usepackage{float}

\usepackage{rotating}
\usepackage{amsmath,amssymb,graphicx,slashed,xcolor,cancel}
\usepackage{array,booktabs,colortbl,multirow,tablefootnote}
\usepackage{colortbl,colordvi,xcolor,color}
\usepackage{caption}
\usepackage{subcaption}
\usepackage{eurosym}

\def\lsim{\mathrel{\rlap{\lower3pt\hbox{\hskip0pt$\sim$}}
   \raise1pt\hbox{$<$}}}         
\def\gsim{\mathrel{\rlap{\lower4pt\hbox{\hskip1pt$\sim$}}
   \raise1pt\hbox{$>$}}}         

\usepackage{relsize}
\usepackage{amsmath}
\usepackage{todonotes}

\mathchardef\mhy="2D   

\definecolor{bk1}{RGB}{0,0,153}

\newcommand{\ecrit}{{\mathcal E}_\textrm{cr}}
\newcommand{\elaser}{{\mathcal E}_\textrm{L}}
\newcommand{\lambdabar}{{\mkern0.75mu\mathchar '26\mkern -9.75mu\lambda}}

\title{Letter of Intent for the LUXE Experiment}

\author[1]{H. Abramowicz}
\author[2]{M.~Altarelli}
\author[3]{R.~A{\ss}mann}
\author[3]{T.~Behnke}
\author[1]{Y.~Benhammou}
\author[3]{O.~Borysov}
\author[4]{M.~Borysova}
\author[3]{R.~Brinkmann}
\author[3]{F.~Burkart}
\author[3]{K.~B{\"u}{\ss}er}
\author[5]{O.~Davidi}
\author[3]{W.~Decking}
\author[6]{N.~Elkina}
\author[6]{H.~Harsh}
\author[7]{A.~Hartin}
\author[3]{I.~Hartl}
\author[3,8]{B.~Heinemann}
\author[9]{T.~Heinzl}
\author[5]{N.~Tal~Hod}
\author[3]{M.~Hoffmann}
\author[9]{A.~Ilderton}
\author[9]{B.~King}
\author[1]{A.~Levy}
\author[3]{J.~List}
\author[10]{A.~R.~Maier}
\author[3]{E.~Negodin}
\author[5]{G.~Perez}
\author[1]{I.~Pomerantz}
\author[3]{A.~Ringwald}
\author[6]{C.~R\"odel}
\author[3]{M.~Saimpert}
\author[6]{F.~Salgado}
\author[11]{G.~Sarri}
\author[5]{I.~Savoray}
\author[6]{T.~Teter}
\author[7]{M.~Wing}
\author[6,11,12]{M. Zepf}

\affil[1]{Tel Aviv University, Tel Aviv, 6997801, Israel}
\affil[2]{Max Planck Institute for Structure and Dynamics of Matter, Hamburg, 22761, Germany}
\affil[3]{Deutsches Elektronen-Synchrotron (DESY), Hamburg, 22607, Germany}
\affil[4]{Institute for Nuclear Research NASU (KINR), Kiew, 03680, Ukraine}
\affil[5]{Weizmann Institute of Science, Rehovot, 7610001, Israel}
\affil[6]{Helmholtz Institut Jena, Jena, 07743, Germany}
\affil[7]{University College London, London, WC1E 6BT, UK}
\affil[8]{Albert-Ludwigs-Universit{\"a}t Freiburg, Freiburg, 79104, Germany}
\affil[9]{University of Plymouth, Plymouth, Devon, PL4 8AA, UK}
\affil[10]{Universit{\"a}t Hamburg, Hamburg, 20148, Germany}
\affil[11]{Queens University Belfast, Belfast BT7 1NN, UK}
\affil[12]{Friedrich Schiller Universit{\"a}t Jena, Jena, 07743, Germany}

\begin{abstract}
This Letter of Intent describes LUXE (Laser Und XFEL Experiment), an experiment that aims to use the high-quality and high-energy electron beam of the European XFEL and a powerful laser. The scientific objective of the experiment is to study quantum electrodynamics processes in the regime of strong fields. High-energy electrons, accelerated by the European XFEL linear accelerator, and high-energy photons, produced via Bremsstrahlung of those beam electrons, colliding with a laser beam shall experience an electric field up to three times larger than the Schwinger critical field (the field at which the vacuum itself is expected to become unstable and spark with spontaneous creation of electron – positron pairs) and access a new regime of quantum physics. The processes to be investigated, which include nonlinear Compton scattering and nonlinear Breit-Wheeler pair production, are relevant to a variety of phenomena in Nature, e.g. in the areas of astrophysics and collider physics and complement recent results in atomic physics. The setup requires in particular the extraction of a minute fraction of the electron bunches from the European XFEL accelerator, the installation of a powerful laser with sophisticated diagnostics, and an array of precision detectors optimised to measure electrons, positrons and photons. Physics sensitivity projections based on simulations are also provided.
\end{abstract}

\begin{document}
\maketitle

\thispagestyle{empty}
\newpage
\tableofcontents

\newpage
\section{Executive Summary}
This Letter of Intent details the plans of a collaboration, comprising leading research institutions and universities in Europe and in Israel, to perform breakthrough experiments in the largely unexplored territory of strong-field quantum electrodynamics (SFQED), exploiting the high quality electron beam of the European XFEL. 

\subsection*{Scientific case}
Quantum electrodynamics (QED) has been outstandingly successful in its predictions in the perturbation regime, reflecting the small value of the fine-structure constant, and applicable to the vast majority of experiments performed so far. Nonetheless, it was predicted \cite{Heisenberg:1935qt}
in the 1930s that in a region of space in which the electric field is strong enough to accelerate an electron to an energy of order $\simeq 0.5$ MeV, equivalent to its rest energy, over the distance of the (reduced) electron Compton wavelength ($1/{m_{e}}\simeq 3.86 \times 10^{-11}$~cm, where $m_e$ is the electron rest mass)\footnote{Natural units are assumed throughout this document: $c=\hbar=\epsilon_0=1$}, perturbation approaches cannot be applied any more and novel phenomena occur; among these are the production of electron – positron pairs by the field-induced tunnelling out of the vacuum, and the \textit{scattering of light by light}, signalling the onset of strong non-linearities in the optical properties of vacuum \cite{Schwinger:1951}.  The field in question is called the \textit{Schwinger field} and amounts to $|\ecrit| = m_e^2/e = 1.32 \times 10^{16}$~V/cm, where $e$ is the charge of a positron. It is at present impossible to achieve a static DC field of such magnitude. Progress in high-power lasers and focusing, however, allows production of fields at optical frequency with \textit{rms} values of the laboratory electric field only a few orders of magnitude short of the Schwinger field. As was shown by pioneering experiments at SLAC in Stanford in the 1990s \cite{Bula:1996st,Burke:1997ew}, it is nonetheless possible to access the region of non-perturbative behaviour in the collision of a high energy electron beam (typically tens of GeV) with the laser photons (typically of order eV). One way to account for this is to recall that in the rest frame of the electrons the electric field of the laser (apart from angular factors of order $1$) is larger by a factor $\gamma_e = E_e/m_{e}$ , where $E_e$ is the electron energy in the laboratory frame of reference. For electron energies exceeding $5$~GeV this factor exceeds $10^{4}$, 
and, with current laser technology, it is possible for electrons to see a field, in   their  rest frame, comparable and even exceeding the Schwinger field. A similar non-perturbative regime is accessed also in collisions of GeV energy photons with the laser photons.

The Stanford experiments in the late 1990s  entered a regime of nonlinear QED, but did not reach the Schwinger critical field, as the optical lasers available at that time were about three orders of magnitude less intense than those available today. Nonetheless they could explore non-linear Compton scattering:
\begin{equation}
e^{-} + n \gamma_L \rightarrow  e^{-} \gamma
\end{equation}
and the Breit--Wheeler process following from further collisions of the nonlinear Compton $\gamma$s with the laser photons:

\begin{equation}
e^{-} + n \gamma_L \rightarrow  e^{-} \gamma \; \; \; \mbox{and} \; \; \; \; \gamma + n \gamma_L \rightarrow  e^{+}  e^{-},
\end{equation}

where $n$ is the number of laser photons, $\gamma_L$, participating in the process.

The observed rates of production for these processes were proportional to higher-than-linear powers of the laser intensity.
The purpose of the LUXE project is to enter far deeper into the non-perturbative region by using the much more powerful lasers available today, together with the high energy electrons from the European XFEL linear accelerator (LINAC).
The dimensionless parameter characterising the intensity of the laser field, $\xi$, is defined as:
\begin{equation}
\xi = \frac{e \elaser} {m_e\omega_{L}}=\frac{m_e\elaser}{\omega_L\ecrit}
\end{equation}
where $\elaser$ is the \textit{rms} electric field of the laser and $\omega_{L}$ its frequency. The region $\xi\ll 1$ corresponds to the perturbative regime; as $\xi$ approaches unity, more and more terms, eventually all of them, need to be retained in the perturbation expansion; and when $\xi>1$ the expansion breaks down completely. Another dimensionless parameter, the \textit{quantum parameter} $\chi_{e}$, accounts for the quantum nonlinear effects in head-on collisions of an electron of energy $\gamma_e m_{e}$  with the laser photons:
\begin{equation}
\chi_{e} =  (2 \gamma_e \frac{ \omega_{L}}{m_{e}}) \xi = 2 \gamma_e \frac{\elaser}{\ecrit} 
\end{equation}
 corresponding to the ratio of the laser \textit{rms} field, in the rest frame of the electrons, to the critical field.

In the Stanford E144 experiment, values of order $\xi \simeq 0.3$ – $0.6$ were attained, still within the perturbative regime, but with observable nonlinear effects in the laser electric field; whereas in the LUXE experiment values up to $\xi\sim 16$ are expected; as well as values up to $\chi_{e} \approx 3$.
The primary motivation of the LUXE experiment is therefore to explore QED in a new regime of very intense electromagnetic fields. The aim is to go beyond the regime in which production rates follow a power of the laser field intensity ($\xi^{2n}$), to enter the region where all orders are equally important and no series truncation can be justified; and to proceed further to higher $\xi$, approaching the asymptotic regime where,  according to Schwinger's prediction, a non-analytic behaviour of production rates as a function of the laser field is expected. There is also a potential that new physics contributions related to new scalar fields (axion or Higgs) could be observed. 
These experiments would be a landmark for the whole subject of strong-field physics, that is also of interest to a variety of disciplines, ranging from astrophysics and cosmology to accelerator physics, and complementing nonrelativistic results in atomic, molecular and condensed matter physics \cite{Brabec:2008}.

\subsection*{Experimental setup}

The European XFEL is designed to run with energies up to $E_{e} = 17.5$~GeV, and contains trains of 2700 electron bunches, each of up to $6 \times 10^{9}$ electrons, that pass at a rate of $10$~Hz. Normally, the European XFEL runs at $E_{e} = 14$~GeV and with $1.5 \times 10^{9}$ electrons per bunch. In this document, $E_{e} = 17.5$~GeV and $1.5 \times 10^{9}$ electrons per bunch is assumed throughout unless otherwise stated. 

The lasers envisaged for this experiment range between $30$ TW and $300$ TW, and are expected to have a repetition rate of $1$~Hz. One electron bunch per bunch train is extracted, yielding a rate of 1 Hz that collides with the laser and an additional 9 Hz that are used for the measurements of beam backgrounds.
Because of the finite response time of extraction kickers, five bunches out of 2700 are expected to be affected, leading to a loss of less than 0.2\% of the beam available for X-ray experiments, which appears tolerable.

\begin{figure}[htbp]
\centering
\includegraphics
[width=\textwidth]{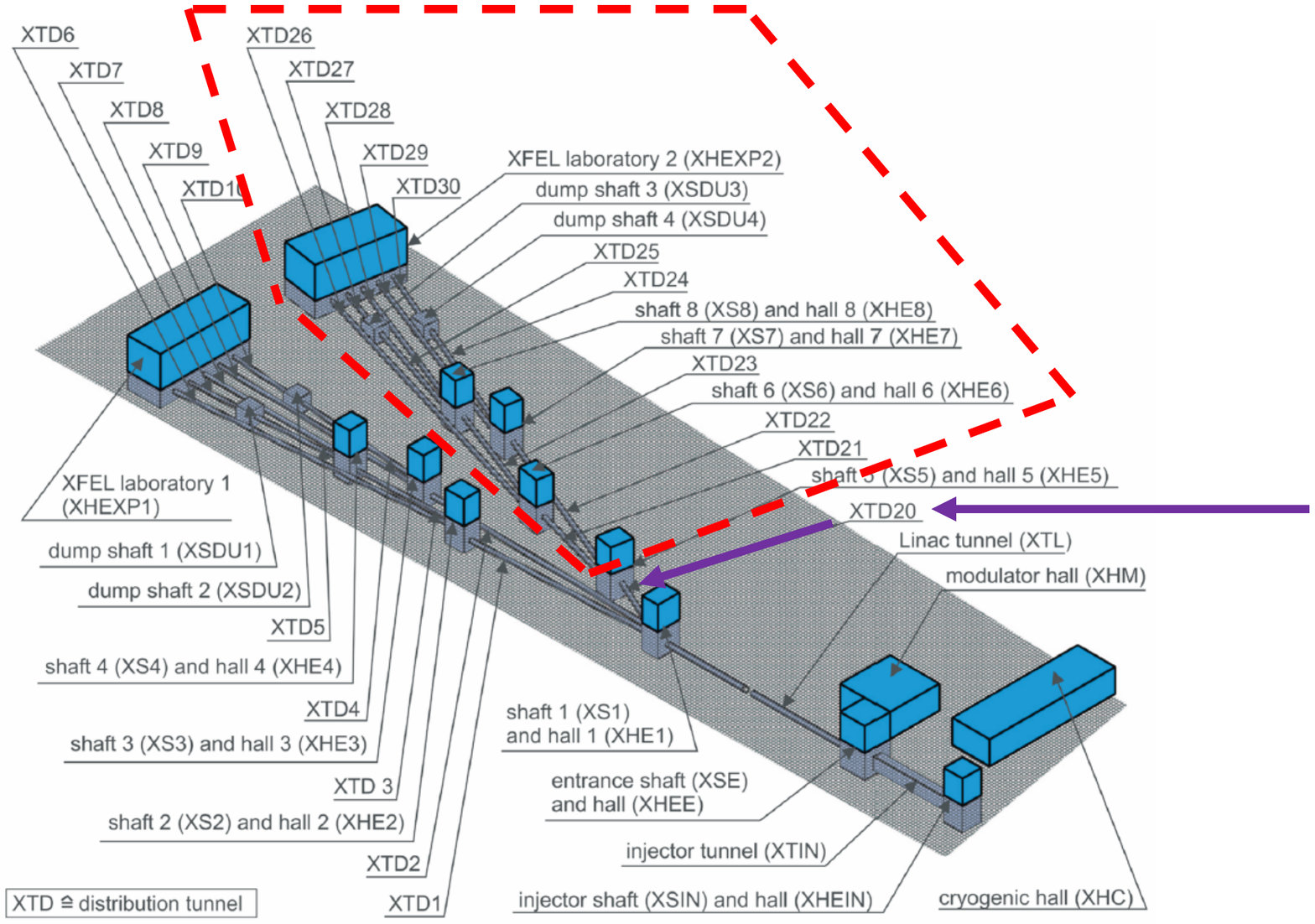} 
\caption{Schematic layout of European XFEL tunnels and buildings. The fan of tunnels inside the red dashed lines is a possible future development, not built yet. The purple arrows point to the annex of the XS1 shaft where the LUXE experiment is foreseen to be installed.} 
\label{fig:tunnel_layout} 
\end{figure}

The location of the bunch extraction upstream of the whole system of undulators, X-ray transport lines and experimental stations, and the steering of the bunches for the LUXE experiment in the presently unused tunnel branch (see Fig.~\ref{fig:tunnel_layout}), foreseen for the later addition of a second fan of undulator tunnels, underline the complete transparency of LUXE assembly and later operation to the X-ray experiments.

\begin{figure}[ht!]
    \centering
    \includegraphics[width=0.4\textwidth]{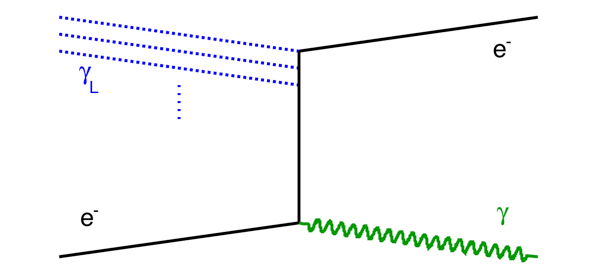}
    \includegraphics[width=0.4\textwidth]{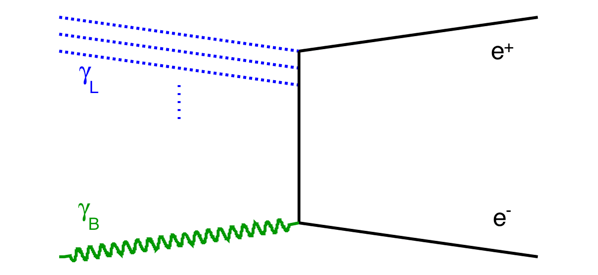}
    \caption{Illustrative diagrams for the dominant processes in the $e$--laser and $\gamma_B$--laser setup.} 
    \label{fig:feynmangraphs}
\end{figure}

The basic idea of the experiment is to collide either the electron bunches directly with the high-power, tight focus laser beam or to send the electrons through a tungsten foil converter, in order to generate photons ($\gamma_B$) by Bremsstrahlung and then collide these with the laser beam, see Fig.~\ref{fig:feynmangraphs}.  

\begin{figure}[ht!]
    \centering
    \includegraphics[width=0.95\textwidth]{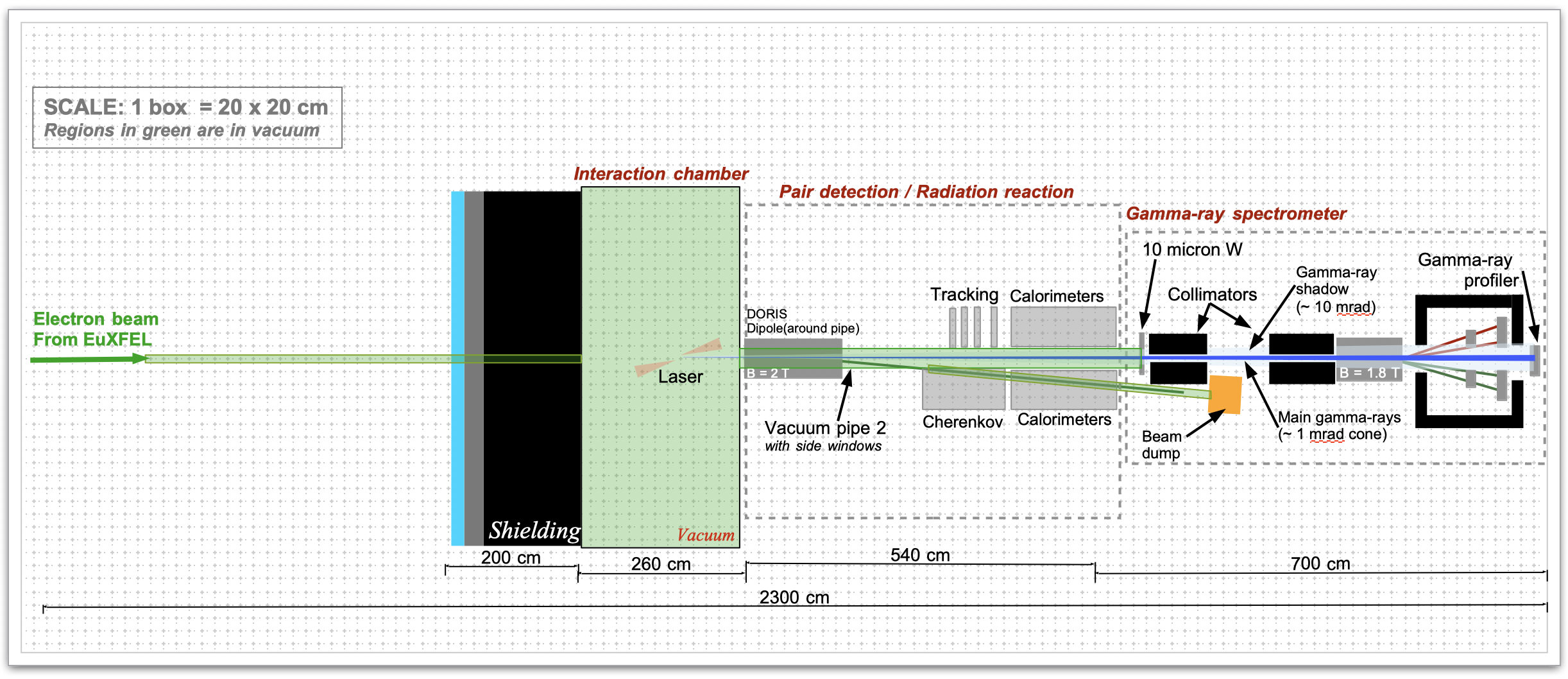}
    \includegraphics[width=0.95\textwidth]{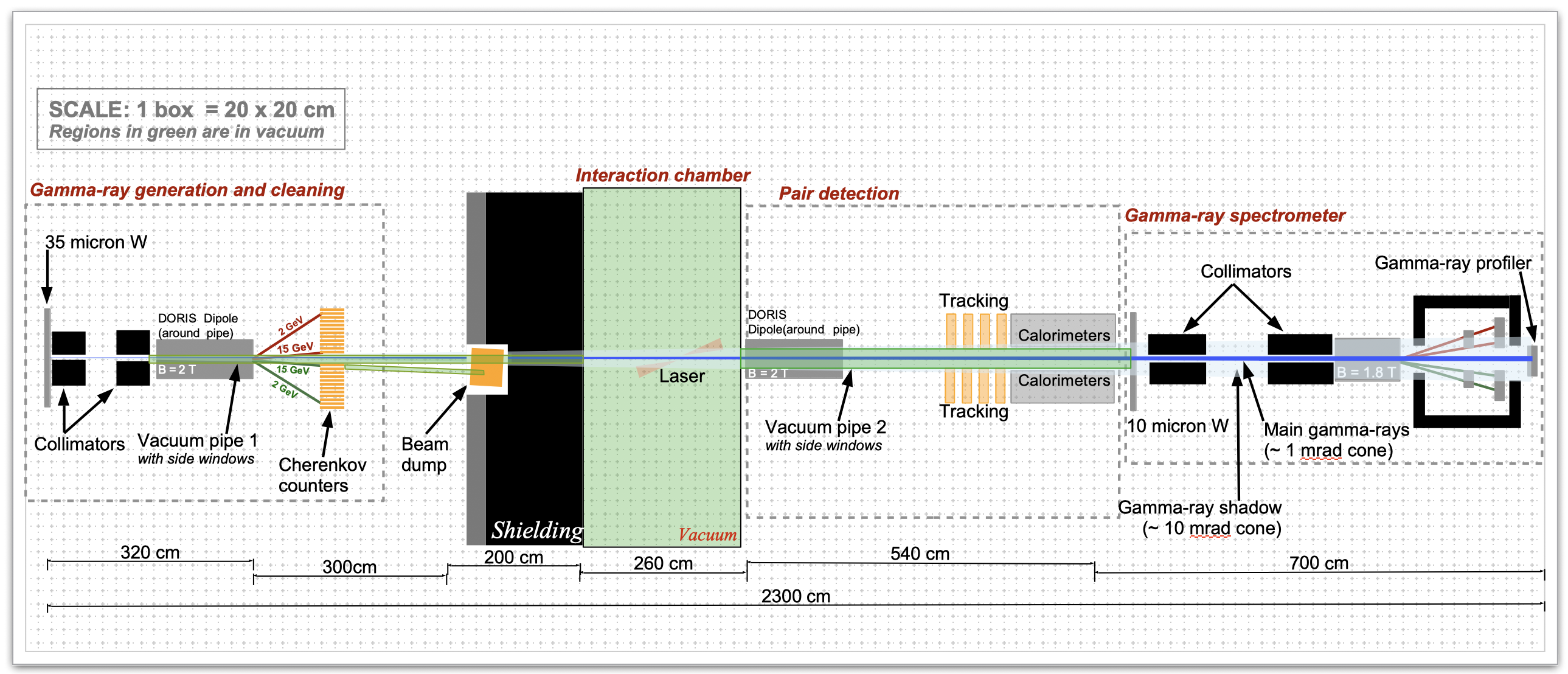}
    \caption{Sketch of the experimental setup for the $e$--laser setup (top) and the $\gamma_B$--laser setup (bottom). The electron beam comes from the left. In both setups a shielding is deployed before the laser-beam interaction point (IP). A dipole magnet and a set of detectors are shown behind the IP. In the $\gamma_B$--laser setup a converter and additional detectors are shown before the shielding. The location of the beam dump is different for the two setups.} 
    \label{fig:introsetup}
\end{figure}

A sketch of the experiment layout is shown in Fig. \ref{fig:introsetup}. 

The basic elements of the apparatus are:
\begin{enumerate}
    \item An electron extraction and beam transport system, starting at the end of the LINAC (see Fig.~\ref{fig:tunnel_layout}) just upstream of the XTD2 and XTD1 tunnels, and running in the presently unused annex to the XS1 shaft (built as a forerunner for a possible future second undulator tunnel fan). A deflection system, composed of a kicker and 4 septa magnets will guide the selected bunch into the XS1 annex. In the electron–laser collision mode, the electrons reach the interaction point $21$~m downstream from the last magnet triplet, and end up in a dump a further $12$~m downstream, after deflection in a dipole that is also used to separate the photons, electrons and positrons generated in the non-linear Compton process, Eq.(1), or in the process, $e^{-} + n \gamma_L  \rightarrow  e^{-} e^{+}  e^{-}$. 
    
In the $\gamma_B$-laser collision mode, the electrons are deflected into a dump upstream from the interaction point, just after the tungsten foil converter that is used as $\gamma_B$ source. 
\item 
A high power laser with  $1$ Hz repetition rate, $30$~fs pulses and an energy per pulse (after compression) ranging from about $1$ J up to $9$ J. The former case corresponds to a power of $30$~TW, the latter to a final power of $300$~TW. The corresponding intensities in a $(8 \times 8) \mu$m${^2}$ focal spot FWHM, assuming standard losses in the optics and transport, amount to approximately $1.6 \times 10^{19}$  W/cm$^{2}$ and $1.6 \times 10^{20}$ W/cm$^{2}$ respectively. A reduction in the size of the laser focus, for the electron–laser setup, to $(3 \times 3) \mu$m${^2}$ implies intensities of about $1.1 \times 10^{21}$ W/cm$^{2}$ for the 300~TW laser.
In the initial phase of experiment focusing will not be pushed to the limits, to allow an easier and more accurate control and reproducibility of intensity and other parameters of the laser pulses, and to give an earlier start to full testing and calibration of other apparatus components. The smaller laser system also requires less physical space and is easier to accommodate with the existing infrastructure. Transport, focusing optics and diagnostics will be designed, built and installed, whenever possible, with the highest power in mind. 

\item A set of detectors designed to detect and measure the electrons, positrons and photons produced. The goal of these detectors is to measure the numbers of particles of the different species, and their energy spectra. A system of silicon tracking detectors, calorimeters and Cerenkov detectors is envisaged. They must have a high granularity so that the number of particles can be counted reliably, even at high intensity, and must have a linear response over a wide dynamic range. In addition, monitoring and diagnostics of the incoming beams is needed and of course a data acquisition and control system.
\end{enumerate}

Typically data taking will take about one month where the laser intensity in focus is changed every few days and measurements are performed at different intensities. In the first year this is done for the electron–laser setup and in the second year for the photon–laser setup.~\footnote{Since the single-bunch beam dump and the detector configuration needs to be changed, it is planned to change during regular end-of-year shutdowns between the two running modes.} This is expected to yield a statistical precision better than 5\% for each measurements point. The systematic uncertainties are expected to be at a similar level. 

The realisation of this project relies critically on expertise in very different areas of physics: high-power lasers, accelerators and particle detectors. The experiment can only be designed and built in synergy between experts from these three areas.

\section{Scientific Goals}
\label{sec:sciencecase}

\subsection{Introduction}
Quantum electrodynamics (QED) describes the interaction of electrically charged particles with electromagnetic  fields. It is often thought of as the most accurate description
of nature and impacts many fields of research in physics. 
Its predictions, for observables accessible by ordinary perturbation theory in the fine-structure constant $\alpha = e^2/(4\pi)\approx 1/137$, where $e$ is the electron  charge magnitude, have been verified up to very high accuracy.  
The most impressive comparison between data and theory is the anomalous magnetic moment of the electron ($g_e-2$): the measured value~\cite{Odom:2006zz,Gabrielse:2006gg} agrees with the theoretical prediction, performed to five loop order ($\alpha^5$) to more than 10 significant digits (see \cite{Volkov:2017xaq} and references therein). For the muon 
$g_\mu-2$, however, there is a $3-4\sigma$ discrepancy~\cite{Blum:2013xva,Jegerlehner:2018zrj} between theory and experiment which could hint at new physics breaking lepton universality, and where a new experiment is now running at FNAL to clarify the situation~\cite{Grange:2015fou}. 

Clearly, the extraordinary precision achieved in perturbative QED has been possible because the  fine-structure constant is very small: $\alpha \ll 1$. However, QED is neither defined, nor exhausted, by just its perturbation series. 
There is QED physics, that is not, and cannot be, captured by a low-order expansion in the QED coupling $\alpha$. A famous example is spontaneous $e^+e^-$ pair production in a static electric field $\mathcal E$, whose rate $\Gamma$ is predicted to be non-perturbative in $e$: 
$\Gamma \propto \exp[-\pi  m_e^2/(e |\mathcal E| )]$, where $m_e$ is the electron mass. However, this awaits experimental verification since the rate is exponentially small for static electric fields smaller than the 
{\it Schwinger critical field}, 
\begin{equation}
\ecrit\equiv 
\frac{m_e^2}{e}\approx 1.3\times 10^{16} \ {\rm \frac{V}{cm}}\ .
\end{equation}
The non-perturbative regime of QED still awaits experimental investigation. 

High field strengths are thought to be present in Nature on the surface of magnetars (strongly-magnetised neutron stars)~\cite{Kouveliotou:1998ze}. They are also expected to be present in future linear high energy $e^+e^-$ colliders~\cite{Bell:1987rw,Blankenbecler:1988te} and in atoms with an atomic number $Z>137$~\cite{pomeranchuk}. Moreover, strong fields - albeit at a scale defined by the ionization potentials - are also of interest in atomic and molecular physics, and the community is striving to achieve them. 

\begin{figure}[ht]
\centering
\includegraphics[scale=0.25]{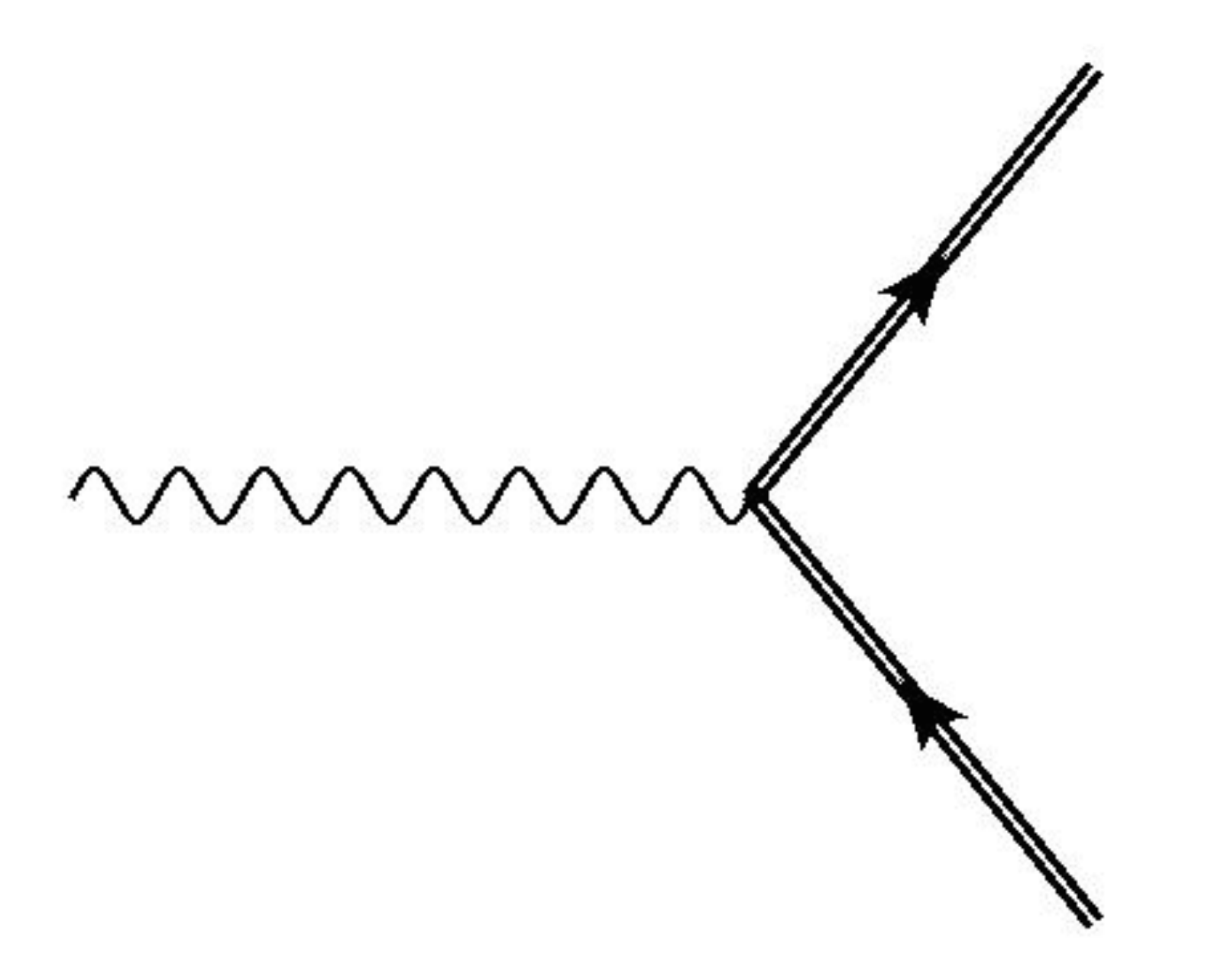} \hspace{2cm}
\includegraphics[scale=0.25]{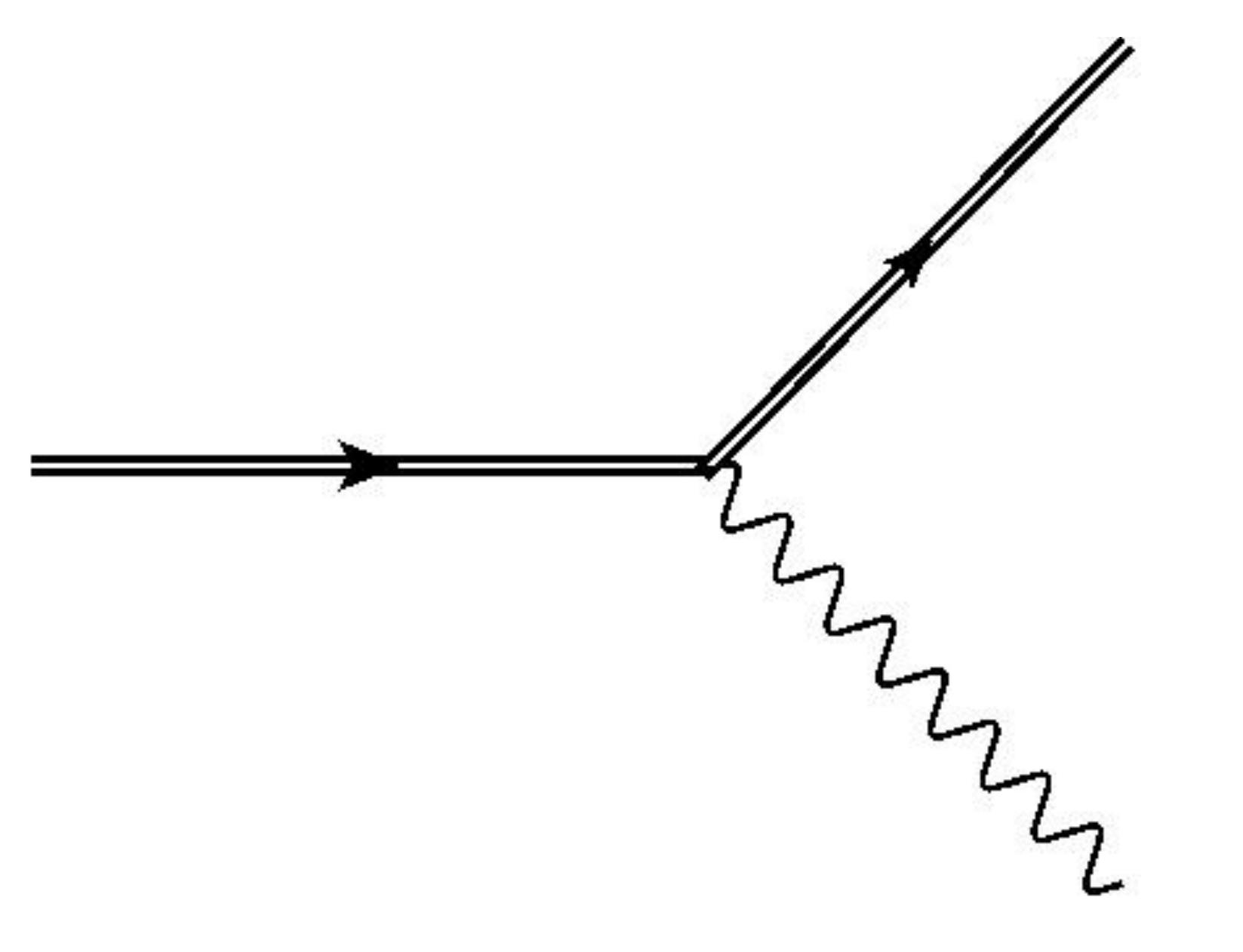}
\caption{Diagrams for processes occurring in the interaction of high energy photons and electrons with intense optical laser beams. Time proceeds from left to right. 
Double lines denote laser dressed (Volkov) electrons (if arrow in positive time direction) and positrons (if arrow in negative time direction). Left: 
Nonlinear pair production (laser stimulated photon decay into an $e^+e^-$ pair - the ``nonlinear Breit--Wheeler'' process). Right: 
Nonlinear photon emission (the ``nonlinear Compton scattering'' process).  } 
\label{fig:NLC.BW} 
\end{figure}

A viable and timely way to probe the realm of non-perturbative QED is via the interaction of high energy photon and electron beams with intense optical laser beams, notably via  non-linear Breit-Wheeler pair production (cf. Fig. \ref{fig:NLC.BW} (left)) and non-linear Compton scattering  (cf. Fig. \ref{fig:NLC.BW} (right)), respectively.
This relies on the fact that, in a laser field $\elaser$ with frequency $\omega_L$, the strength or intensity parameter is \begin{equation}
\label{eq:xi}
\xi^2=4\pi\alpha
\left[\frac{\elaser}{\omega_L m_e}\right]^2=
\left[\frac{m_e}{\omega_L}\frac{\elaser}{\ecrit}\right]^2.
\end{equation}
It then follows that the probability of a net absorption/emission of $n$ photons from the laser beam is proportional to $\xi^{2n} \sim \alpha^n$ which is consistent with perturbative QED vertex counting \cite{McDonald:1986zz}. On the other hand, $\xi^{2}$ is proportional to the number density of photons, $n_L$, namely $\xi^2 = 4 \pi \alpha \lambdabar_L \lambdabar_{C}^{2} n_L$, where $\lambdabar_L$ and $\lambdabar_C$ denote the reduced laser and Compton wavelengths, respectively. If $\xi \gsim 1$, it is no longer sufficient to calculate a single $n$-vertex process of magnitude $\xi^{2n}$ as \emph{all} of these, hence each $n$, can contribute with comparable weight. In other words, the perturbation series in interactions between background and charges can no longer be truncated. Thus, one enters a regime of non-perturbative physics at small coupling, requiring solutions valid ``to all orders'' in perturbation theory (in $\xi^2 \sim \alpha$). In very intense fields $\xi\gg1$, one is even in a regime where processes are described by a dependency on $\alpha$ for which no perturbation series can be defined. 
All of this leads to a growing field of research called \emph{high-intensity QED} or \emph{strong-field QED} \cite{Heinzl:2008an} defining a ``high-intensity frontier'', where the flux of photons is so large that physics is again ``all-order'' and the standard perturbative methods of QED cannot be applied.
 With the advance of high-power laser technology based on chirped pulse amplification~\cite{Strickland:1985gxr} (for which the 2018 Nobel Prize in physics was awarded), a whole new field of experimental high-intensity laser interactions with fundamental quantum systems has been opened~\cite{Dunne:2008kc,DiPiazza:2011tq}. This has led to recent successes at e.g. the UK Astra-Gemini laser facility~\cite{Cole:2017zca,Poder:2018ifi} and triggered the next-generation of high-intensity lasers such as the European Extreme Light Infrastructure (ELI)~\cite{Zamfir:2014msa}, which are currently in the construction phase.

\begin{figure}[ht]
\centering
\includegraphics[scale=0.25]{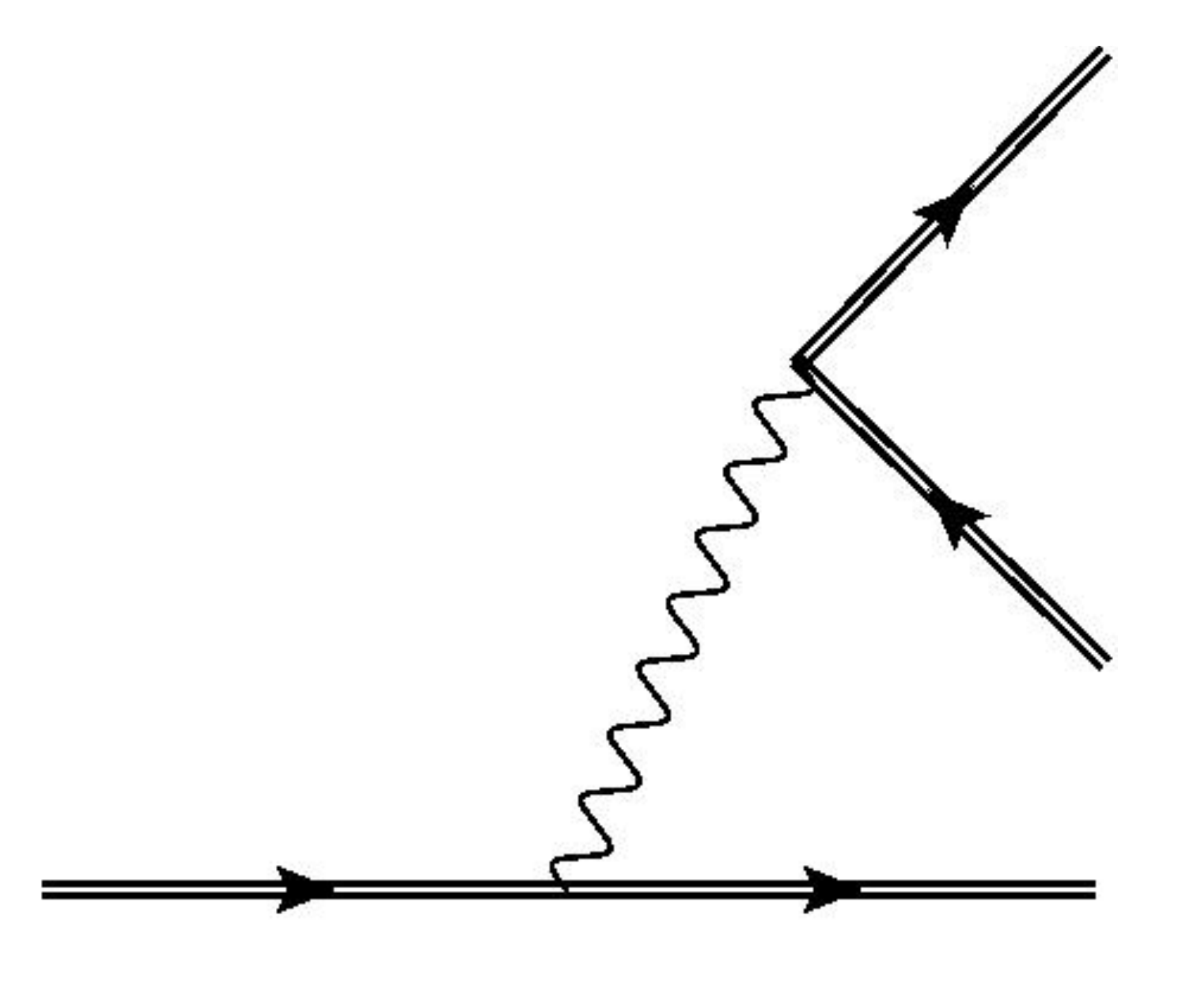} 
\caption{Leading order diagram for the laser stimulated trident process. Cutting through the internal photon line produces the processes of Fig.~\ref{fig:NLC.BW}. The process can either proceed in a quasi-instantaneous single step ("one-step trident") or via sequential subprocesses shown in Fig.~\ref{fig:NLC.BW} ("two-step").} 
\label{fig:trident} 
\end{figure}

In the 1990s, the  E144 experiment~\cite{Bula:1996st,Burke:1997ew}, was realised at SLAC. It used the 46.6 GeV electron beam from the Stanford Linear Collider (SLC) and a green laser with 1 TW power. It was a landmark experiment that aimed to reach the QED small-coupling non-perturbative regime by studying electron-laser interactions. 
E144 observed nonlinear Compton scattering in the emission of a photon resulting from electron laser interactions with up to $n=4$ laser photons \cite{Bula:1996st} through processes of the form $e + n\, \gamma_L\, \to e + \gamma$. These events were generated experimentally by bringing 46.6\,GeV electrons into collision with a laser with intensity parameter $\xi  \simeq 0.4$. In a second step, the backscattered laser photons of energy 29.2\,GeV produced electron positron pairs via nonlinear Breit-Wheeler processes, $\gamma + n\, \gamma_L \to e^+ e^-$. Overcoming the energy-momentum threshold for pair-creation required the net absorption of at least $n=5$ laser photons and this was experimentally confirmed \cite{Burke:1997ew}. This process is illustrated in Fig.~\ref{fig:trident}. 

In principle, this two-step process (nonlinear Compton followed by nonlinear Breit-Wheeler) can also proceed in a single step through the exchange of a \emph{virtual} photon via the (laser dressed) trident process, its naming stemming from the fact that an initial electron ends up in a three-particle final state (an outgoing electron and a pair, see Fig.~\ref{fig:trident}). Using the Weizs\"acker–Williams approximation, E144 estimated the direct trident pair production rate to be suppressed by three orders of magnitude compared to the two-step process measured (see Fig.~5 in \cite{Burke:1997ew}), although it is now known that the Weizs\"acker–Williams approximation can fail when the laser field is sufficiently intense~\cite{King:2013osa}.

However, E144 fell short of reaching the all-order regime \cite{reiss1,Hu:2010ye}, $\xi\gtrsim 1$. In recent years, laser technology has advanced significantly, so that now lasers exceeding 100 TW power are commonplace. By combining such lasers with the European XFEL electron beam, the all-order region can be reached~\cite{Hartin:2018sha}, so that, for the first time, the non-perturbative regime can be explored experimentally. LUXE will play a pivotal role in this. Since solving non-perturbative physics problems is generally very challenging, any experimental input will assist theory development and testing its predictions. 

For pair creation to proceed, the centre-of-mass energy must overcome the pair rest mass threshold. This is quantified by the \emph{lightfront parameter}:
\begin{equation}
\eta_i \equiv \frac{p_i\cdot k}{m_{e}^2}=(1+\cos\theta)\frac{\omega_L E_i}{m_e^2},
\end{equation}
for particle $i=e,\gamma$ and laser photon momenta $p_i$ and $k$ respectively\footnote{Here,  $\theta$ is the angle of collision between the particle $i$ and the laser propagation direction ($\theta=0$ corresponds to a ``head-on'' collision).}. When the electron lightfront parameter $\eta_{e}$ fulfills $\eta_{e} > 2(1+\xi^{2})/n$ for integer $n$, pair creation can occur if a net of at least $n$ photons are absorbed from the laser field. Whether pair-creation is probable to occur, is quantifiable with the \emph{quantum parameter}:
\begin{equation}
\label{eq:chii}
 \chi_i \equiv \xi \eta_i=(1+\cos\theta)\frac{E_i}{m_e}\frac{\elaser}{\ecrit}.
\end{equation}
$\chi_{e}$ corresponds to the laser electric field in the rest frame of the electron, in units of the Schwinger field $\ecrit$ and so, when $\chi_{i} \approx O(1)$, pair-creation is more likely to occur, whereas, if $\chi_{i}\ll1$, pair-creation is exponentially suppressed. In the E144 experiment values of $\chi_{e}\approx 0.3$ were reached, and more recently, values close to $\chi_{e} \approx 1$ have been achieved in static crystalline fields in an experiment at CERN using a tertiary electron beam with energy of around 180~GeV~\cite{Wistisen:2017pgr}. However, the deeply quantum regime of $\chi_{e}\gtrsim 1$ has yet to be probed experimentally.

\begin{table}[]
    \centering
    \begin{tabular}{c|c|c|c|c|c|c|c}
     Phase& \multicolumn{2}{c|}{Power [TW]} & focus & Intensity& $\xi$ & \multicolumn{2}{c}{$\chi_e$ for $E_e$/GeV}\\
      & nominal & actual & FWHM [$\mu$m] &  [$10^{19}$~W/cm$^2$] & & $17.5$ & $14.0$\\\hline
         A& 30 & 12 & 8 & 1.6 & 2 & 0.41 & 0.32\\
         B& 300 & 120 & 8 & 16 & 6.2 & 1.3 & 1.0\\
         C& 300 & 120 & 3 & 110 & 16 & 3.3 & 2.6\\
    \end{tabular}
    \caption{Parameters of the LUXE experiments for the electron-laser mode. Given are the nominal and actual laser power (accounting for losses of $\sim 60\%$), the FWHM of the focus area, and the corresponding laser intensity, $\xi$, and $\chi_e$ values for two electron beam energies. Further details on the parameters are given in Sec.~\ref{sec:laser}.}
    \label{tab:xivalues}
\end{table}

LUXE will have the capability of running in two different modes. I) In the first mode, it will combine a high-energy and well-characterised electron beam with an intense optical laser pulse. It will produce both high particle energies, $\eta_{i} \approx 0.2$,  and high intensities, $\xi = 0.5 \ldots 16$, thereby accessing the largest values of the quantum parameter,  $\chi_i$ in excess of $3$, yet seen in experiment (see Table~\ref{tab:xivalues}).
It will have the discovery potential for  high-intensity QED processes such as one-step trident scattering. It will also measure the non-linear Compton and the two-step trident process in a new regime. II) In the second mode, LUXE will collide the electron beam with a $\gamma$ converter (a thin tungsten foil), to generate a beam of high-energy Bremsstrahlung photons ($\gamma_B$) which can be collided directly with the high-power laser pulse, comprising photons of frequency $\omega_L$. The main reaction of interest in this mode is non-linear Breit-Wheeler. With all these measurements, the LUXE experiment will advance the field significantly compared to previous experiments, notably SLAC E144 \cite{Bamber:1999zt}. 

\subsection{Details on Physics Processes}
The primary motivation of the LUXE experiment is to explore QED in a new regime of very intense electromagnetic fields. With the measurement of high-intensity QED processes, the goal is to characterise and study the transition to all-order, non-perturbative small coupling physics predicted in quantum field theory (QFT) \cite{McLerran:1999wf}.
The aim is to go beyond the perturbative \emph{multiphoton} regime in which production rates follow a power law dependency, $\xi^{2n}$ (for laser field intensity $\xi$ and threshold number of photons $n$), and enter the region where all orders of interaction between the laser and the probe particle can be equally important and no series truncation can be justified; and to proceed further to even higher $\xi$  , approaching the asymptotic regime where, according to Schwinger’s prediction, a non-analytic behaviour of production rates as a function of the laser field is expected. More generally, this is the regime predicted by quantum field theory of non-perturbative physics at small coupling, which also occurs in other contexts. All three processes discussed above (nonlinear Compton, Breit-Wheeler and trident) are predicted to exhibit relativistic quantum signatures of the transition to the non-perturbative regime when $\xi\gg 1$ and $\chi_e\gtrsim 1$.

Quantum Chromo Dynamics (QCD), the theory of the strong interaction, is a non-abelian QFT based  on the SU(3) group, and is so far the only QFT tested in the non-perturbative regime. However, the nature of this non-perturbative regime is different to QED, in the sense that the coupling constant, $\alpha_s$, is large. A phenomenon related to the non-pertubativity at small coupling that arises in high-intensity QED, is the saturation expected at low values of Bjorken-$x$ where the number of gluons in the proton increases drastically but is expected to reach saturation at some point. The large number of interacting gluons forming a glass condensate correspond to the large number of interacting laser photons forming a classical field. This has been searched for at HERA and is part of the goal of many future electron-hadron colliders~\cite{Accardi:2012qut,Abada:2019lih,Caldwell:2017zfz}.

With the LUXE experiment, the non-perturbative regime for a QFT based on the U(1) group will be tested, which could lead to new insights into QFT itself and computational techniques used to model QFT. Since there is no standard method that tackles all non-perturbative problems, experimental input is vital in assisting theory development. Any advances in solving non-pertubative physics in QED can therefore potentially positively influence the solution of similar, challenging, physics in other contexts.

The measurement of all three processes will allow the transition from the perturbative to the non-perturbative regime to be examined. Below, we introduce the phenomenology of Breit-Wheeler pair-creation to illustrate this transition.

In the low-$\xi$ multiphoton regime, a good approximation for the rate is given by calculating only the process in which the minimum number $n_{\ast}$ of laser photons is involved. This minimum corresponds to a pair created at rest, which in a circularly-polarised monochromatic laser background fulfills $n_{\ast} = \lceil 2(1+\xi^{2})/\eta_e\rceil$. The probability then follows a power law of the form $P\propto \xi^{2n_{\ast}}$. LUXE is expected to see this power-law scaling by varying the laser intensity and measuring event rates as a function of $\xi$. When the laser intensity is further increased, so that $\xi\gtrsim 1$, the perturbative hierarchy is disrupted and processes of higher order can in principle contribute more than lower-order processes. Thus an all-order calculation must be performed where contributions from every order are summed. An example is the asymptotic result for the probability when $\xi_{\gamma}\gg1$, $\chi<1$, \cite{ritus85}
\begin{equation}
    \label{eq:rate}
    P\propto \chi_\gamma e^{-8/(3\chi_\gamma)}.
\end{equation}
Since $\chi_\gamma \propto e \propto \sqrt{\alpha}$, the probability cannot be expanded pertubatively and hence represents an experimental signature for non-perturbative effects. It is also seen that this probability is small when $\chi_\gamma\ll 1$ and only becomes appreciable when $\chi_\gamma\sim 1$
The LUXE experiment aims to probe the values of $\xi$ and $\chi_\gamma$ relevant to observe this transition from perturbative QED $P\propto \xi^{2n}$ to non-perturbative QED, such as demonstrated by the scaling $P\propto \chi_\gamma e^{-8/(3\chi_\gamma)}$~\cite{Hartin:2018sha}. 

\begin{figure}[ht]
    \centering
    \includegraphics[width=0.7\textwidth]{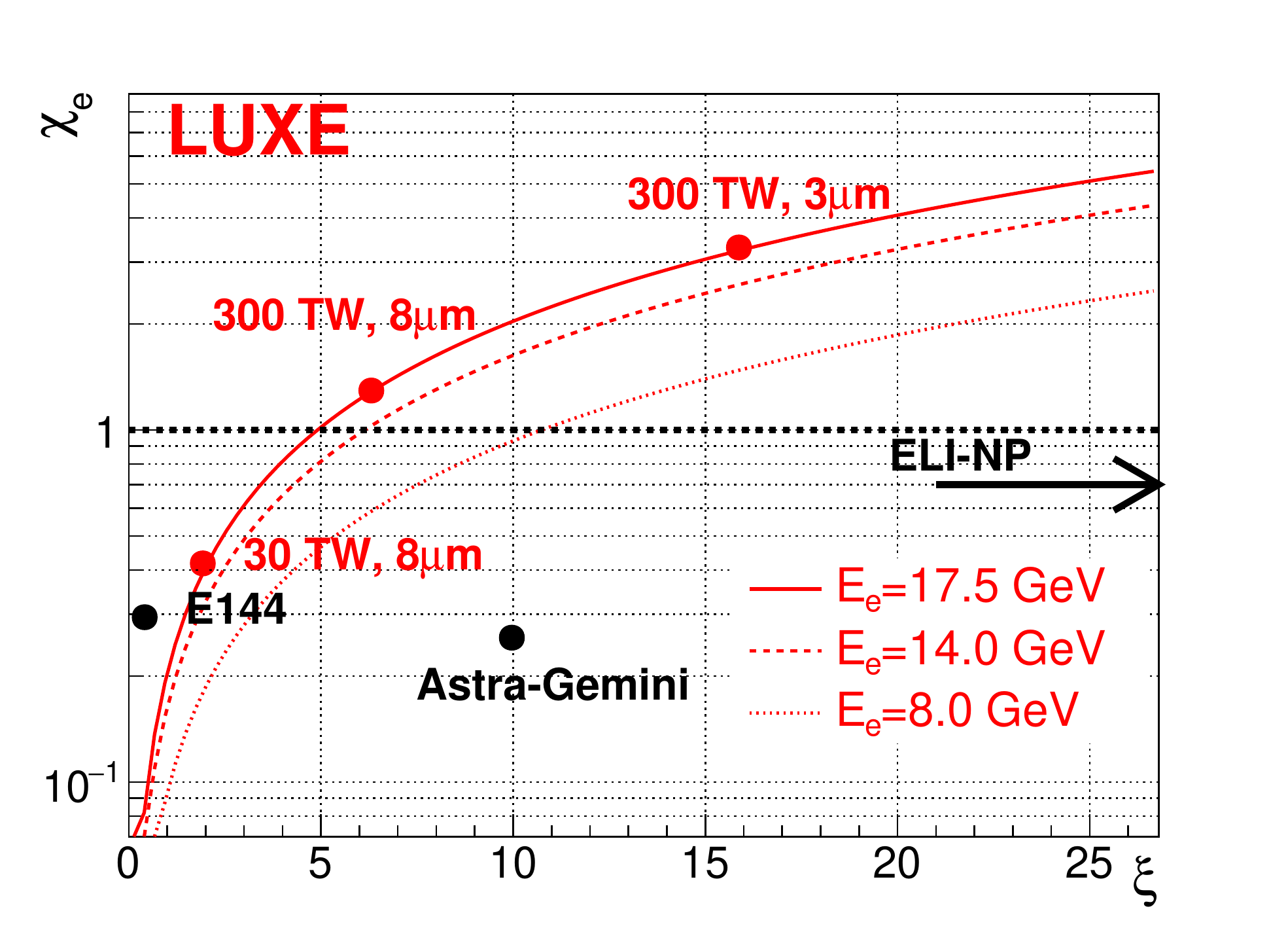}
    \caption{
    The $\chi$ and $\xi$ parameter space accessible by various experiments. The three red lines show the parameters accessible to LUXE for three electron beam energies. Indicated are the parameters corresponding to the foreseen laser configurations of LUXE: a 30 TW laser focused to a FWHM of $8$~$\mu$m, or a 300 TW laser focused to either a FWHM of $8$~$\mu$m or $3$~$\mu$m. For details on the laser assumptions, including the assumed light yield in focus, see Sec.~\ref{sec:laser}. Also shown are the parameters reached by the E144 and ASTRA-GEMINI experiments. ELI-NP will reach $\xi$ values beyond those shown on this plot. \label{fig:chivxi}
    }
\end{figure}

Figure~\ref{fig:chivxi} shows $\chi_e$ as a function of $\xi$ for the LUXE experiment for electron beam energies of $17.5$, $14.0$ and $8.0$~GeV. It is seen that $\chi_e \gtrsim 1$ is exceeded for the 300~TW laser. Lower values of $\xi$ and $\chi$ are accessed by defocussing the laser and thereby reducing the laser intensity. For a 30~TW laser, the $\chi$ value corresponds approximately to that reached by E144 but the $\xi$ value is now about $3\times$ larger for the nominal energy of $17.5$~GeV.

By measuring the rate of the pair production process versus $\xi$ and $\chi$, at low $\xi$ and $\chi$ it can be tested if it follows the expected power law ($\propto \xi^{2n}$) but as $\xi$ and $\chi$ increase it should depart from the power-law due to non-perturbative effects and follow the exponential behaviour of Eq.~\ref{eq:rate}. The goal is to measure this behaviour and to extract the exponent with a precision of about 10\%.  

Nonlinear Compton scattering is the primary process occurring in the $e+n\gamma_{L}$ setup, and we explain here some basic phenomenology. If the laser pulse can be modelled as quasi-monochromatic and circularly-polarised, the magnitude of the electromagnetic field can be considered to be quasi-constant. A free electron in this
wave undergoes a circular motion with angular frequency
$\omega_L$ in the plane transverse to the direction of propagation
of the laser field. The electron's transverse momentum is of the order $P_\perp\sim\xi m$, so that the total energy of the electron is given by (recalling $c=1$)
\begin{equation}
    E^2  =  m^2+P_\perp^2+P_\parallel^2
     \sim (1+\xi^2) m^2+P_{\parallel}^2\, ,
\end{equation}
where $P_\parallel$ is the component of the electron's momentum
parallel to the direction of propagation of the wave. In that sense, it is often said that the electron behaves as if it had a higher ``effective mass'' $\overline{m}$:
\begin{equation}
    \overline{m}=m\sqrt{1+\xi^2}\,,
\end{equation}
with correspondingly altered ``effective momentum'' \cite{Brown:1964zzb}. This behaviour is identifiable by a shift in the harmonic range for lowest-order Compton scattering (the Klein-Nishina process) called the ``Compton edge'' \cite{Harvey:2009ry}. The effective mass also occurs in the threshold number of photons $n_{\ast}$ required for Breit-Wheeler pair-production, which determines the power-scaling in $\xi$ in the multiphoton regime. In finite laser pulses, the spectral signature of the effective mass can be affected by the pulse envelope \cite{Harvey:2012ie} and can exhibit harmonic resonance substructure \cite{Heinzl:2010vg}. 

\begin{figure}[ht]
    \centering
    \includegraphics[width=0.68\textwidth]{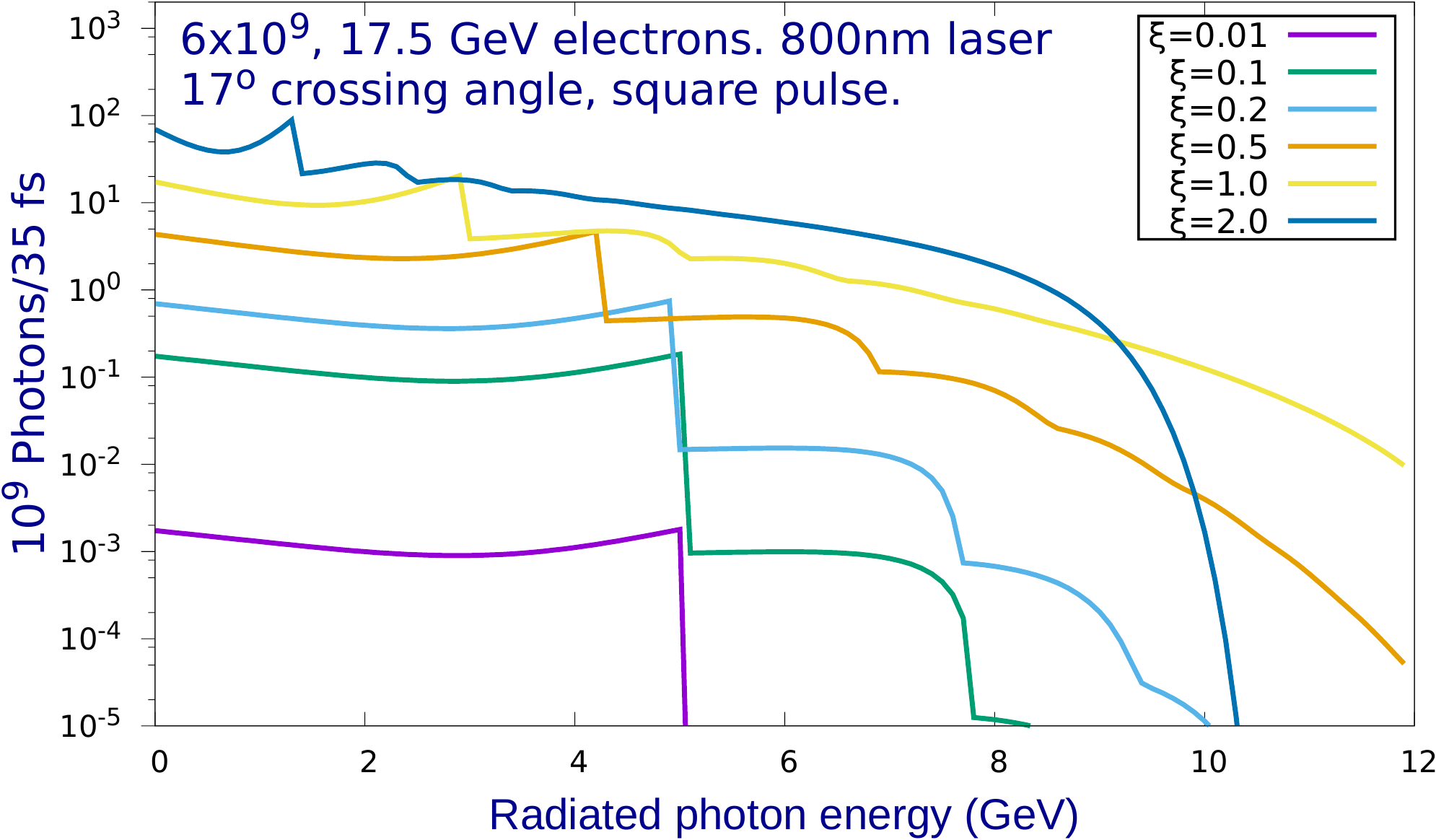}
    \caption{Nonlinear Compton-scattered photon energy spectrum for $6\times 10^9$ electrons 
    in a circularly-polarised quasi-monochromatic background pulse of duration 35~fs. 
    \label{fig:compton}}
\end{figure}

Fig.~\ref{fig:compton} shows examples of photon energy spectra for several values of $\xi$ for $E_e=17.5$~GeV. The different edges on each curve correspond to a  different net number of laser photons participating in the interaction. It is seen that for the different $\xi$ values the position of the Compton edge is also different. For $\xi=0.01$ the first edge is at 5~GeV, and with increasing $\xi$ it is shifted to lower and lower photon energies, roughly scaling as $1/\sqrt{1+\xi^2}$. The 2nd and 3rd edges are also seen at higher values of the photon energy. The overall flux of photons is high, e.g. for $\xi=0.2$ it is about $10^7$ photons for an electron beam containing $1.5\times 10^9$ electrons. And, at values $\xi\gsim 1$ the number of photons exceeds $10^{9}$. This large number makes the counting of photons and the measurement of their energy spectrum challenging. This task is performed by a photon detection system and is discussed in Sec.~\ref{sec:detectors}. 

When the nonlinear Compton scattered photon interacts again in the laser, the two-step trident process occurs if the high-energy photon produced by the Compton scattering interacts with the laser field and an $e^+e^-$ pair is produced. This process was observed by E144~\cite{Burke:1997ew}, and it will also be measured to much higher precision at LUXE and compared to theoretical predictions. 

The two-step trident process, which proceeds via an on-shell photon, depends on the spectrum of nonlinear Compton scattering and the rate of the Breit-Wheeler process. While the two-step trident rate generally increases with the laser intensity, it is limited by the lower energy photons available from the mass shifted Compton spectrum. These competing tendencies result in the rate increasing initially, but then falling sharply at higher $\xi$-values. A similar behaviour has been reported in the constant field case \cite{King:2013zw}, which is relevant to the high-intensity $\xi\gg1$ regime. Optimal two-step trident production, based upon the circularly-polarised monochromatic sub-process rates, lies in the range $2.5<\xi<3.5$ depending on the initial electron energy.

The trident process also receives contributions from a ``one-step'' process, in which the photon exchange is virtual. At the SLAC E144 experiment, the contribution from this process was estimated to be negligible, but this was based on a Weizs\"acker-Williams approximation, rather than an explicit knowledge of the one-step contribution. The one-step process has received significant attention during recent years~\cite{Ilderton:2010wr,Dinu:2017uoj,King:2018ibi,Mackenroth:2018smh}. It is already known for example that the one-step and two-step channels scale differently with pulse shape, and give different final-state momenta distributions of produced pairs. It is also known that in the constant field case, the one-step process can suppress the total probability \cite{ritus72,King:2013osa} which may affect predictions based solely on the two-step process.  The tools are being developed to analyse to what extent the one-step contributes in the LUXE experiment, and what experimental signals it may contribute to. The extent to which the measured positron signal differs from the known two-step signal, will give an experimental determination of the one-step contribution.

The experimental setup proposed here could also be beneficial for studying some forms of physics Beyond the Standard Model (BSM). It has been shown that theories introducing new scalar fields, as in the case of axion-like and Higgs-portal models, would also induce photon non-linearities~\cite{Evans:2018qwy,Flacke:2016szy}.
Such effects could be enhanced in the strong field background generated by a laser.
While non-linear photon dynamics are also expected within the Standard Model (SM), via the Heisenberg-Euler Lagrangian~\cite{Heisenberg:1935qt}, it has yet to be measured.
Furthermore, as was pointed out in Ref.~\cite{Bogorad:2019pbu}, BSM effective photon self-interactions could be sizeable, and possibly even dominate over photon-photon scattering as predicted in the SM.
Thus, the intense fields to be produced in this experiment potentially form an ideal setting to probe non-linear effects related to new physics. In particular, a preliminary study suggests a unique sensitivity to CP violating phenomena, which currently lie in an uncharted territory of BSM physics.

\subsection{Simulation Results}

The projected measurements and uncertainties are evaluated using Monte Carlo (MC) simulation. The MC simulation relies on a calculation similar to those described in Refs.~\cite{Hartin:2018egj,Hartin:2018sha}. Both the $e+n\gamma_L$ and the $\gamma_B+n\gamma_L$ process are generated. For the latter, the flux of $\gamma_B$ used in the simulation is determined using a GEANT4~\cite{Allison:2016lfl,Allison:2006ve,Agostinelli:2002hh} simulation of the electron beam hitting the converter (a $W$ foil, see Sec.~\ref{sec:detectors}). Typically, 10\% of the beam electrons radiate a photon when using a target with $35$~$\mu$m. 

The generated events are then passed through a fast simulation program that determines their path through the experimental setup described in Sec.~\ref{sec:detectors}. Basic simulations of the detectors reflecting their granularity and efficiency are also included. 

The background is estimated using Monte Carlo simulations of the experimental setup based on independent implementations in GEANT and FLUKA~\cite{fluka}. The main concern are secondaries produced when electrons or photons hit material, e.g. the magnets, beam dump, collimators, beam pipe etc. The collimators and shieldings and beam pipe geometry have been optimised to minimise the background: with the current statistics it is estimated that in the region after the IP there is less than 20 positrons from secondary interactions with $E>1$~GeV per $1.5\times 10^9$ beam electrons which enter the detector acceptance. These originate from the beampipe, and it is anticipated that they can be rejected based on precision tracking information. The goal is to optimise the setup such that the background is less than $0.1$ positrons per laser shot.

\begin{figure}[ht]
    \centering
    \includegraphics[width=0.48\textwidth]{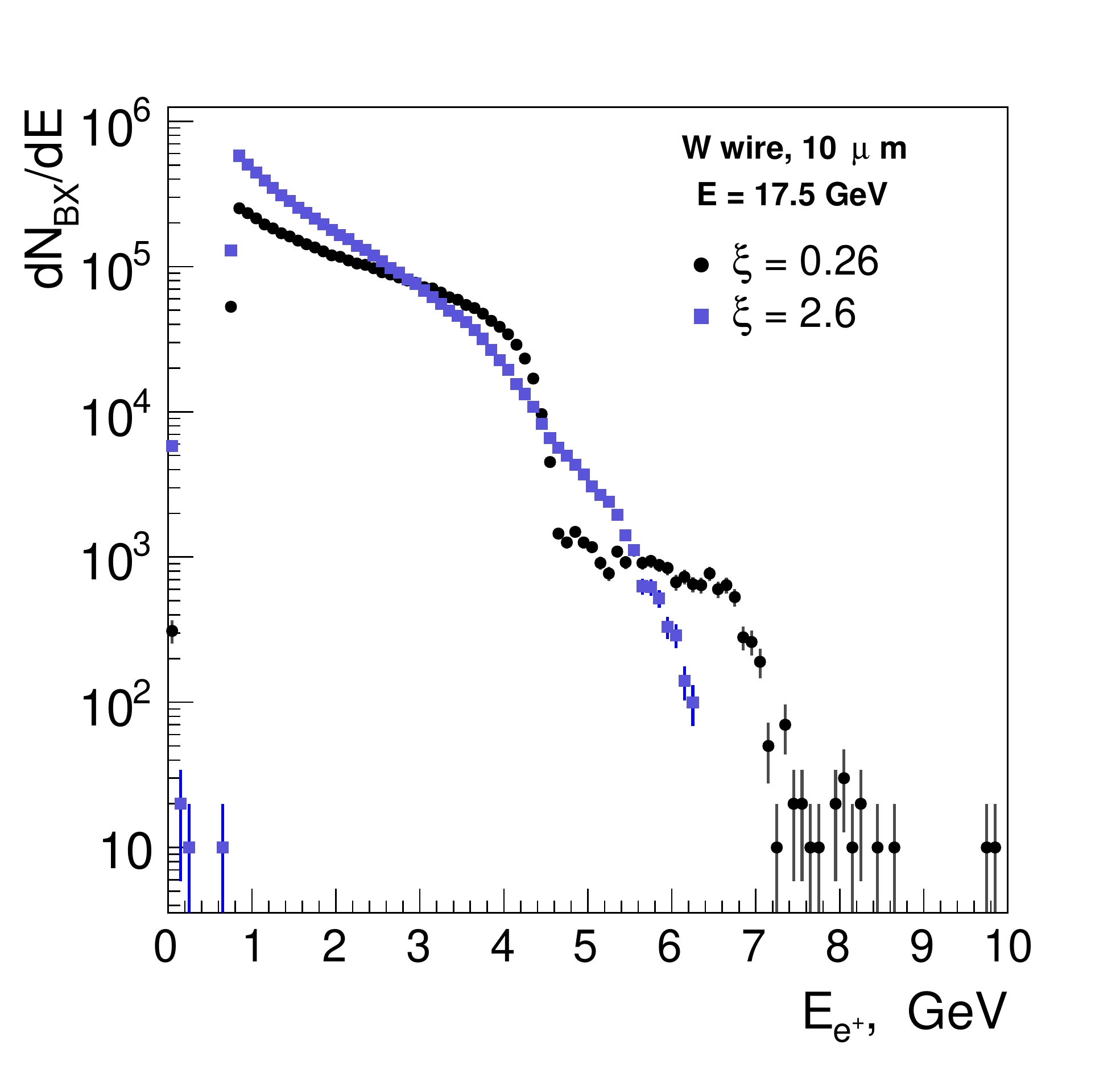}
    \caption{Simulated positron energy spectrum for $\xi=0.26$ and $\xi=2.6$ and a $10$~$\mu$m tungsten wire (see Sec.~\ref{sec:detectors}) and $1.5\times 10^9$ electrons per bunch. 
    \label{fig:comptonpos}}
\end{figure}

For the $e+n\gamma_L$ setup it is expected that both the non-linear Compton and the two-step and one-step trident processes occur. For the study of the non-linear Compton process the photon detection system and the electron detector are most important. In the electron detector, the numbers of electrons can be counted in intervals of electron energy using Cherenkov detectors, see Sec.~\ref{sec:detectors}. For the measurement of the photon spectrum, a fraction of photons are converted to $e^+e^-$ pairs using a thin wire, and these electrons and positrons can be used to determine the number of photons and the photon energy spectrum. Fig.~\ref{fig:comptonpos} shows the positron energy spectrum that is observable in the photon detection system. The Compton edges as expected for the photon (see Fig. \ref{fig:compton}) at low $\xi$ values are observed also in the positron energy spectrum.

\begin{figure}[ht]
    \centering
    \includegraphics[width=0.45\textwidth]{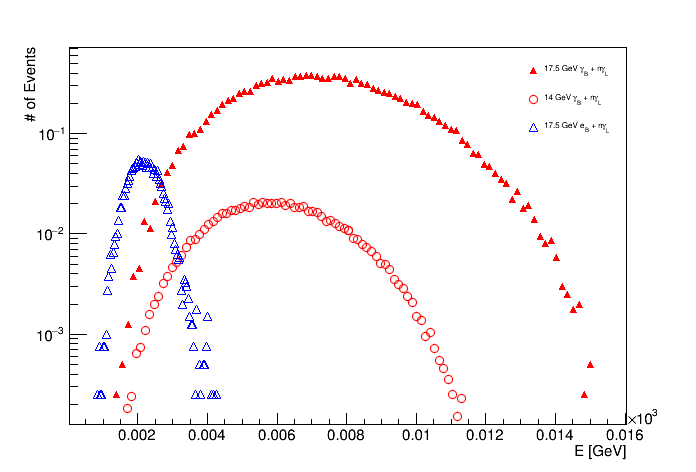}
    \includegraphics[width=0.45\textwidth]{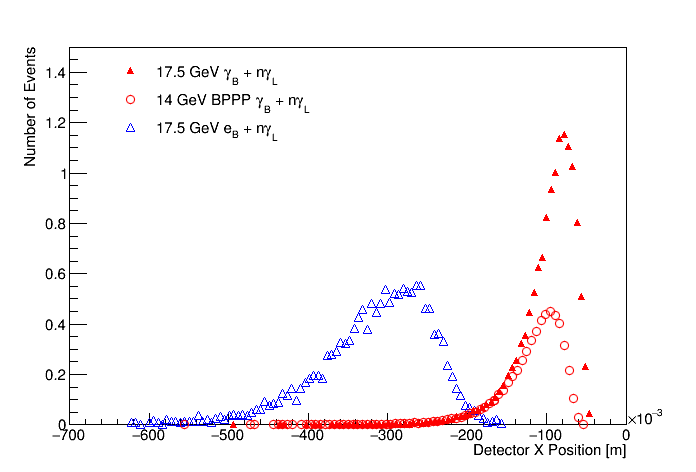}
    \caption{Left: Positron energy spectra for the pair-production (for $E=17.5$~GeV and $14$~GeV) and the two-step trident processes for $\xi=2.6$. Right: Horizontal position of the positron at the detector plane after the magnet for the same samples.
    \label{fig:espec}}
\end{figure}

The primary objective for the $\gamma_B+n\gamma_L$ process is the measurement of the positron rate as a function of the $\xi$. Fig.~\ref{fig:espec} shows the positron energy spectrum for $\xi=2.6$ for $E_e=14$~GeV and $E_e=17.5$~GeV. It is seen that for a beam energy of $17.5 \, GeV$ the spectrum covers the range of $2-14 \, GeV$ and has a maximum near $7 \, GeV$. For a beam energy of $14 \, GeV$, the rate is reduced by over an order of magnitude and is shifted to somewhat lower energies. For the two-step trident process the positron energy spectrum is also shown in Fig.~\ref{fig:espec}. It is seen that the energies are significantly lower. Fig.~\ref{fig:espec} also shows the transverse position of the positron impact at the detector plane. Most of the particles are contained between 4 and 54~cm. Requiring them to be in this $|x|$-range corresponds to an acceptance of 95\% for the two-step trident process and 100\% for the pair--production process for $B=1.4$~T. 

\begin{figure}[ht]
    \centering
    \includegraphics[width=0.8\textwidth]{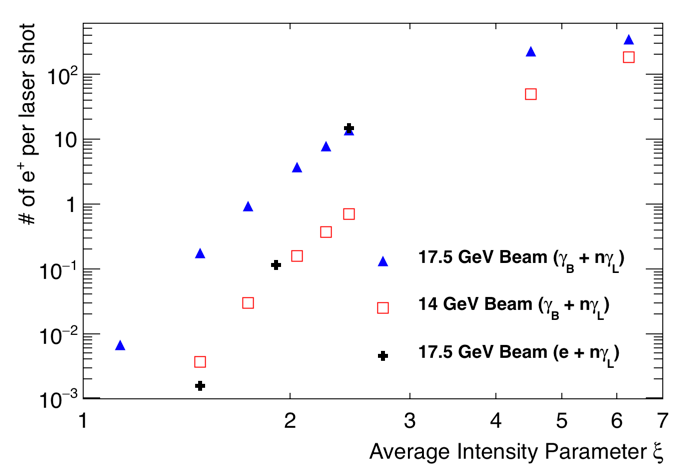}
    \caption{Number of positrons per laser shot versus the mean value of $\xi$ in the $\gamma_B$--laser (for $E_\textrm{beam}=17.5$ and $14.0$~GeV) and the two-step trident process in the $e$--laser setup. No uncertainties are shown. It is expected that the statistical precision achieved will be between about 10\% at low $\xi$ and $\ll 1\%$ at high $\xi$. 
    \label{fig:luxeresults}}
\end{figure}

The expected detected positron rate versus the average value of $\xi$ is shown in Fig.~\ref{fig:luxeresults} for both processes. The average $\xi$ value can be calculated if the shape of the laser pulse and its energy are known (see Sec.~\ref{sec:laser}). It is typically 10\% lower than the peak value for the laser parameters used here.

It is seen that for the pair production process in the $\gamma_B$--laser setup, the number of positrons per laser shot increases from $7\times 10^{-3}$ to $350$ within the $\xi$-range considered for $E_e=17.5$~GeV; for $E_e=14$~GeV the rates are about a factor 70 lower. In both cases the rate increases initially steeply, following a power-law, and then the increase slows due to the non-perturbative effects, so that it now has an exponential form (see previous subsection). 

With the 30~TW laser values up to $\xi=2$ will be measurable, for higher $\xi$ values the 300~TW laser will be needed. For the two-step trident process the rates are also shown for $E_\textrm{beam}=17.5$~GeV. It is about 10 positrons per laser shot at $\xi\approx 2.5$.

In all cases, it is clear that it is desirable that backgrounds are suppressed to levels below 0.1 particles per laser shot, so that measurements can be performed in the entire range of $\xi>1$ with a precision of $\lsim 5\%$. 
For a signal event rate of 0.01 per laser shot and background rate of 0.1 per electron beam crossing, and assuming that the background is measured \textit{in-situ} with a rate of 9~Hz\footnote{If only 1~Hz of background is taken the precision degrades from 10\% to 15\% for a signal rate of 0.01~Hz.}, 24h are needed to measure the signal event rate with a precision of 10\%. For a signal event rate of 0.1 per laser shot the precision is 1.5\% and it improves further for higher values of the signal rate. If the background rate cannot be controlled to better than 1 event per laser shot, the uncertainty degrades to 40\% for 0.01 signal events and to 4\% for 0.1 events per laser shot.

Assuming an up-time of 30\%, the 24h of actual data taking would take about 3 days to collect. For higher laser intensities, the statistical precision would be correspondingly better. Normally, we will want to scan the laser intensity, and take data for about 2-3 days, then readjust the laser intensity and again take data for 2-3 days. With a total of about 10 measurement points and assuming the laser intensity adjustment takes at most one day, the actual measurement programme should take about one month once the experiment is fully commissioned and understood.

\section{Electron Beam Transport and European XFEL Accelerator Aspects}
\label{sec:machine}

In this section a brief summary of the modifications to the European XFEL necessary for the LUXE experiment is given. A more detailed description is given in Ref.~\cite{beamlinecdr}.

The LUXE experiment aims to be fully transparent for European XFEL operation, therefore it is foreseen to use the standard European XFEL beam parameters. An overview of the electron beam parameters for the LUXE experiment is given in Table \ref{tab:epara}. The frequency of the bunch train is 10~Hz. Therefore the rate is limited by the repetition frequency of high quality laser shots, which is only 1~Hz.

\begin{table}[htbp]
\begin{center}
\begin{tabular}{l|c}
Parameter                    & Value \\ \hline
Beam Energy [GeV] & up to 17.5           \\ \hline
Bunch Charge [nC]             & 0.25–1.0             \\ \hline
Number of bunches  & 1                \\ \hline
Repetition Rate [Hz]           & up to 10                \\ \hline
Spotsize at the IP [$\mu$m]        & 5–20               \\ \hline
\end{tabular}
\caption{Electron beam parameters envisaged for the LUXE experiment. At the highest energy, the bunch charge is taken to be 0.25~nC by default for this document. The electron bunch length is 30–50~$\mu$m, much longer than the laser pulse of 9~$\mu$m (30~fs, see sec.\ref{sec:laser}). 
\label{tab:epara}
}
\end{center}
\end{table}

\subsection{Location at the European XFEL}
\begin{figure}[ht]
   \centering
   \includegraphics*[height=4.2cm]{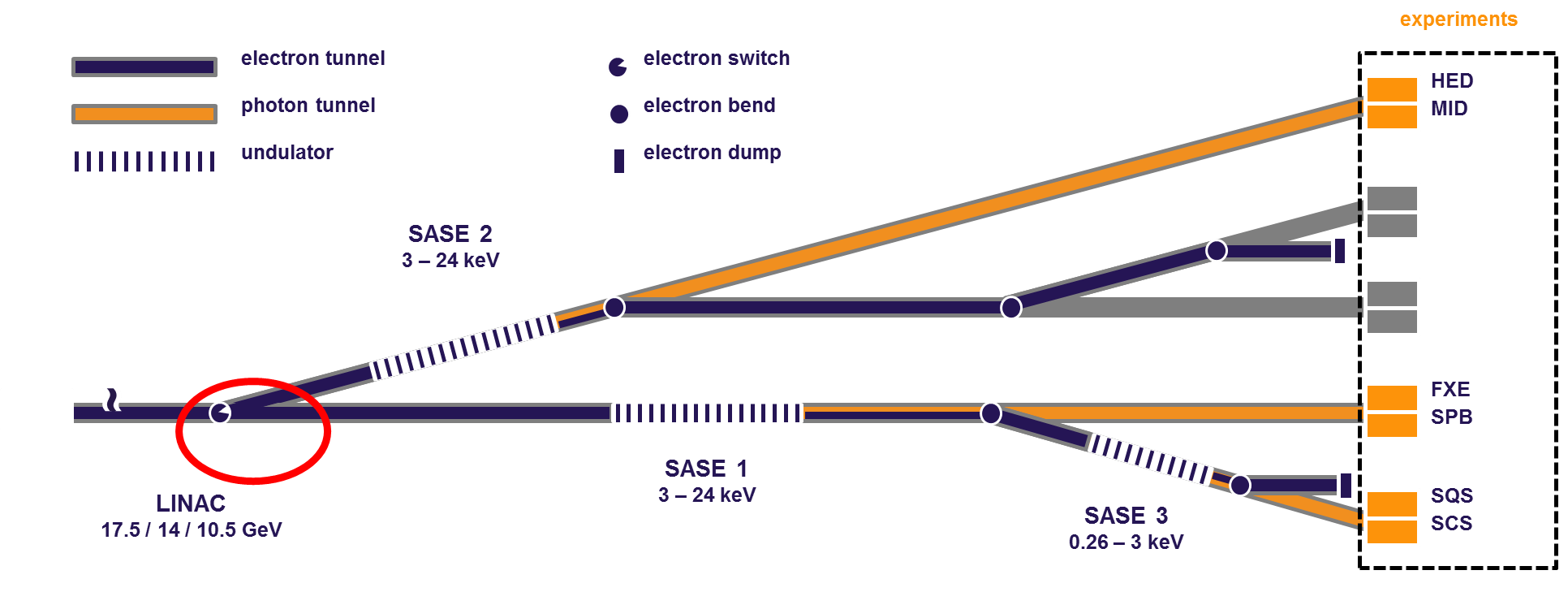}
   \caption{Schematic drawing of the European XFEL fan. The location foreseen of the LUXE experiment is circled in red. The XT1 and XT2 beamlines are those serving the SASE2 and SASE1 undulators, respectively. 
   \label{fig:XFEL_sketch}}
\end{figure}

It is planned to install the LUXE experiment in the XS1 annex which is an appendix of the XS1 shaft, located in Osdorfer Born, that has already been foreseen for a possible extension of the distribution fan of the XFEL. It is about 60~m long, 5.4~m wide and 5~m high. It is at the end of the main LINAC of the European XFEL (XTL). A drawing of the European XFEL is shown in Fig.~\ref{fig:XFEL_sketch}. 

\begin{figure}[ht]
   \centering
   \includegraphics*[height=4.2cm]{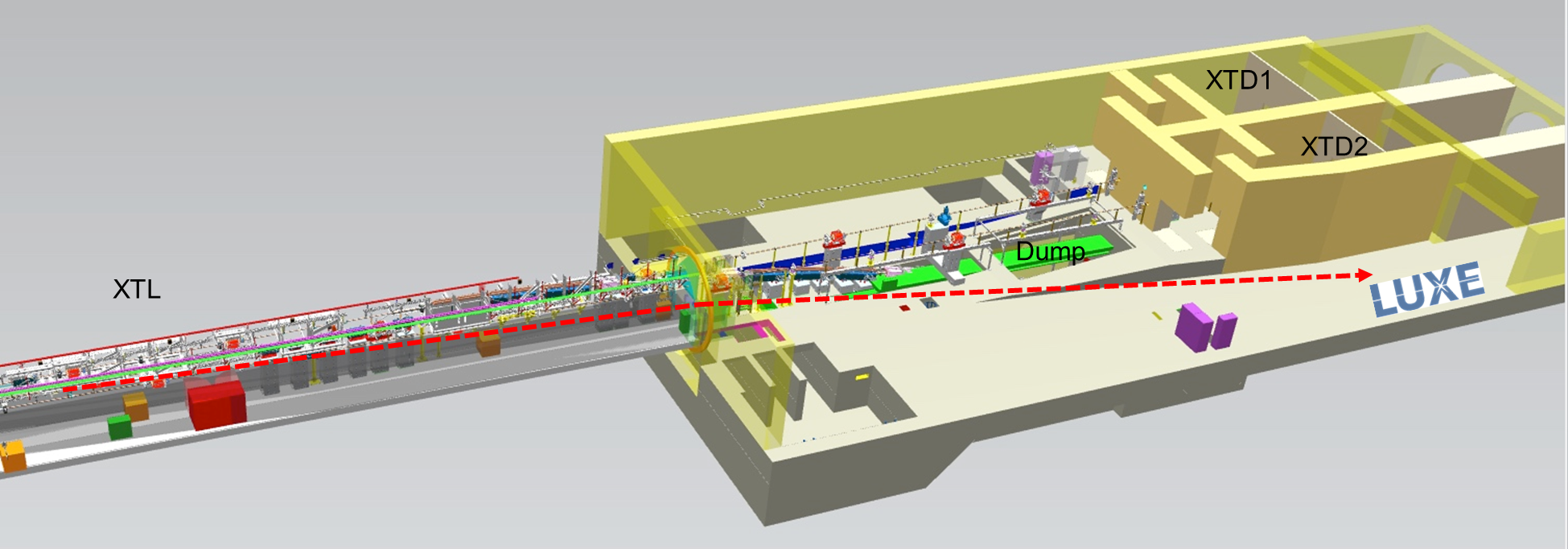}
   \caption{CAD model of the end of the European XFEL accelerator tunnel and the shaft building with the two existing
beamlines XT1 and XT2 to the undulators (SASE1 and SASE2) and the XS1 annex, where the LUXE experiment can be installed. The beam
extraction and the beam line towards the experiment is sketched with the dashed line. }
   \label{fig:LUXE_cad}
\end{figure}

The beam path to the XS1 annex, required for LUXE, and the XTD1 and XTD2 tunnels are also schematically shown in Fig.~\ref{fig:LUXE_cad}. Also shown is the LINAC, XTL. The beam transport to the experiment has to be built up with an additional extraction. This was already foreseen in the design phase of the European XFEL for the possible future extension of the facility. About 40~m of installations in the XTL tunnel are needed to kick out one bunch and steer the beam towards the XS1 annex. Inside the XS1 shaft building additional components are needed to steer and focus the beam to the experiment. The length of this installation is about 50~m. 

\subsection{Lattice Design and Hardware}
\begin{figure}[ht]
   \centering
   \includegraphics*[width=0.8\textwidth]{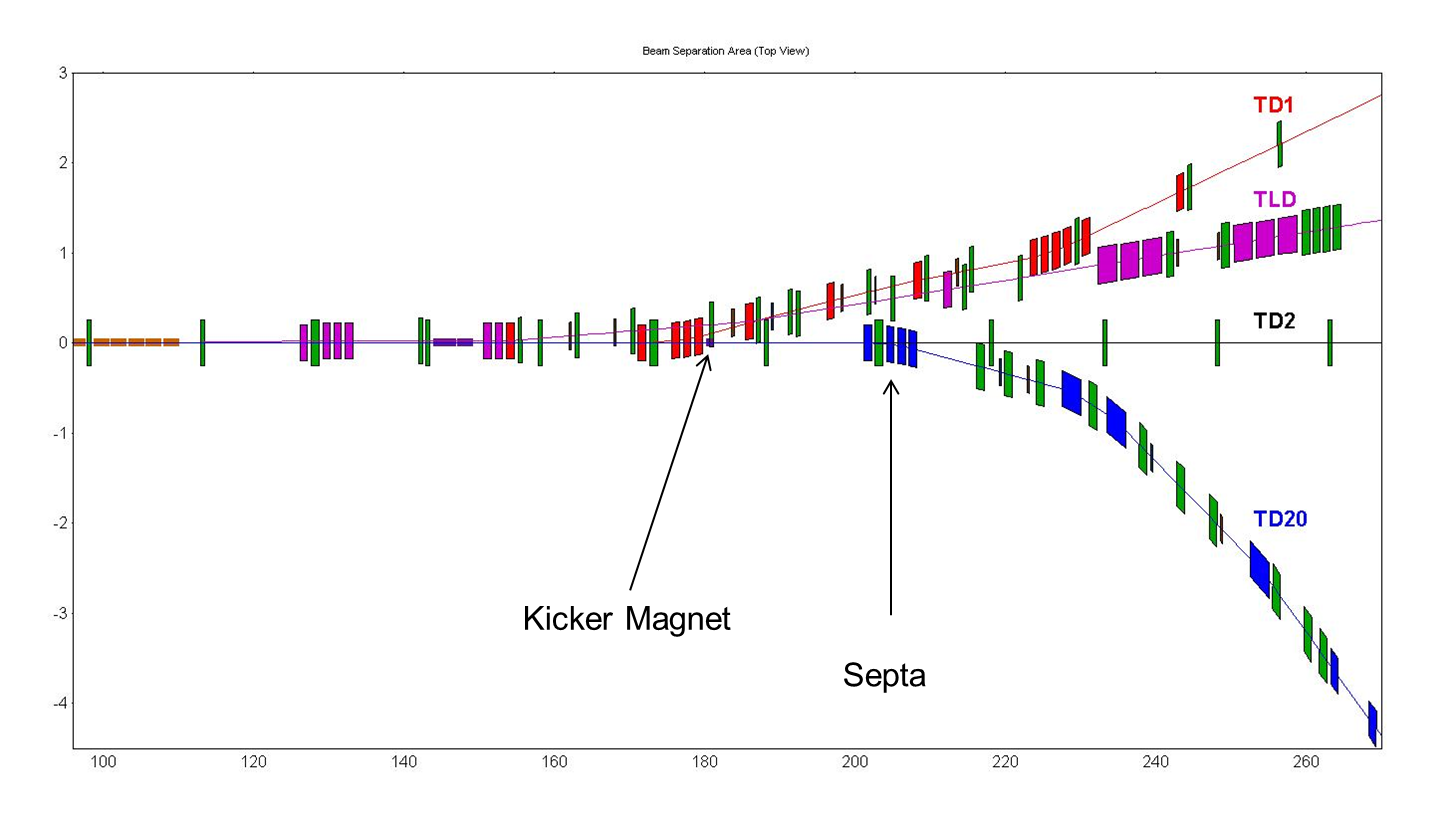}
   \caption{Design of the beamline extraction. The different magnets (kicker, dipole, quadrupole) are shown in different colors. The new kicker magnet which kicks out one bunch towards LUXE is indicated. The horizontal and vertical scales are in units of m. The LUXE experiment starts at the end of the area shown ($z=270$~m). The XS1 building starts at $z\approx 235$ m in this drawing, $30$~m upstream from the Septa indicated. 
   \label{fig:lattice}}
\end{figure}

The design for the beamline is shown in Fig.~\ref{fig:lattice}. The part until the last Septum is termed \textit{beam extraction} and that after the Septa is termed \textit{beam transfer line}. 40~m of the beam transfer line are located inside the XTL tunnel, whereas the rest has to be installed in the XS1 shaft building and the XS1 annex. A combination of fast kicker magnets, septa, dipole and quadrupole magnets is used for the design of the beam extraction and beam transfer. 
In the beam extraction part two kicker magnets and 4 septa are used. In the transfer line,  11 dipole magnets, 10 quadrupoles and 10 corrector dipoles are needed. In addition, an array of diagnostic elements are foreseen.  

For the kicker magnets, a new design will be used~\cite{beamlinecdr} while all other components adopt the designs already used in the European XFEL. Also for power supplies and general infrastructure, e.g. vacuum system, beam instrumentation, water cooling, cabling, safety systems the already installed and tested standard European XFEL components will be used.

\subsection{Installation Procedure}
The installation of the extraction and transfer line requires major construction work both in the XTL tunnel and in the XS1 shaft. Additional infrastructure has to be installed in the XS1 building. While some of this prepatory work can be done in parallel to XFEL operation, the bulk of the work has to be done during shutdown periods. 
In order not to affect European XFEL operation, the installation of the extraction and transfer line is planned to be compatible with regularly scheduled shutdown periods. It should be noted, that during these periods the technical personnel of DESY is fully committed to regular European XFEL maintenance and upgrade work. Thus this additional installation work requires the hiring of additional qualified personnel.

The different phases of the installation are listed in the following:
\begin{itemize}
    \item The \textbf{preparatory phase} consists of design, specification, tender, ordering, fabrication, delivery and testing of all necessary components.
    \item For the \textbf{installation in the XTL tunnel} it is foreseen to pre-assemble the vacuum chambers and septum magnets outside the tunnel (as has been done for the installation of the XTD1 extraction). A total of five weeks are expected to be needed for the installation. 
    
    The installation of the vacuum chamber for the two kicker magnets is considered a minor intervention and will be performed separately in a previous shutdown.
    The installation sequence would result in restoring full functionality of the XTL beamline after a shutdown period. 
    \item The \textbf{installation of the transfer line} concerns all components after the last septum. These components steer the beam towards LUXE. They are in a vacuum system that is separate from the already existing beam lines. Nevertheless the installation of the components in the XTL tunnel (about 30~m) is challenging as they are suspended from the ceiling. The installations in the XS1 shaft building are expected to be straightforward. A total of seven weeks are required for this installation after the last septum. Since the vacuum system of the beam transfer line to LUXE is separate from the rest of the European XFEL, any delays here will only delay LUXE and not pose risks for the European XFEL.
\end{itemize}

\subsection{Beam Dump}
After the electron beam bunch has traversed the tungsten target it needs to be dumped safely. For this purpose a special beam dump has been designed. It has a maximum charge of 1~nC and rate of 10~Hz, resulting in a maximum power of $P=200$~W for a beam energy of $E=20$~GeV which need to be dumped. In addition, shielding is required to ensure that there is no contamination of nearby air, water or rock, people can safely work in the adjacent tunnel, nearby electronics equipment is not damaged by radiation, and, last but not least, that the background from stray particles in the LUXE experiment is low enough so that the foreseen physics analyses can be performed (see Sec.~\ref{sec:sciencecase}). 

\begin{figure}[htbp]
   \centering
   \includegraphics*[width=0.8\textwidth]{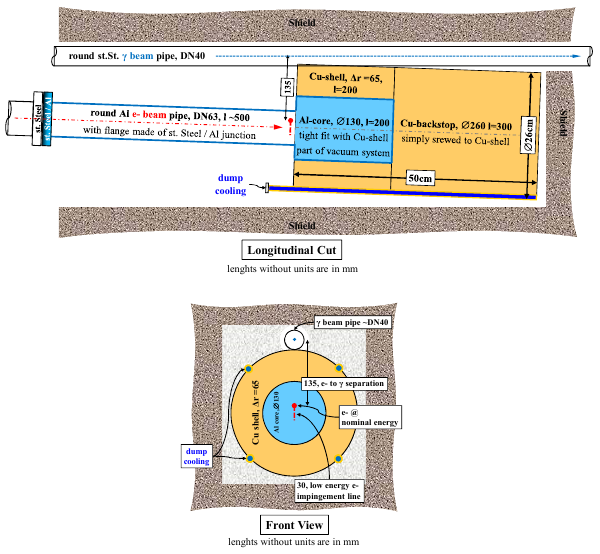}
   \caption{Schematic drawing of the beam dump.
   \label{fig:beamdump}}
\end{figure}

A schematic drawing of the beam dump is shown in Fig.~\ref{fig:beamdump}. It consists of a aluminium core with a diameter of 13~cm and a length of 20~cm, which is complemented by copper surrounding it and also extending its length. The total length is 50~cm, and it is integrated into the shielding before the interaction point. The beam dump is located 3.5~m downstream from the magnet. Water cooling is used to cool it in 4 cooling pipes. 

For the different experimental modes of LUXE the beam dump has to be moved to the required position. In the $\gamma_B$ mode it is expected to be at a distance of at most 20~m from the end of the focusing triplet magnets. This implies that there are about 16~m available for the laser IP and detection system.

\subsection{Risks and Commissioning}
The risks have been estimated in Ref.~\cite{beamlinecdr}. If a delay occurs due to equipment being damaged in either the beam transfer or the transfer line, the shutdown period may need to be prolonged. The work plan foresees to mitigate this risk by using trained personnel, proper planning and good workmanship, and estimates the risk to be below 5\% based on prior experience at the European XFEL. Another risk is that the vacuum cannot be closed due to missing or wrong equipment. For the XTL tunnel, this risk is mitigated by keeping the current system fully available so that one could revert to that system if needed. The likelihood is estimated to be below 5\%; it would cause a delay of two weeks. For the transfer line in the XTD20 tunnel the vacuum is separate and any delays would only affect the LUXE experiment and not the European XFEL at large. The likelihood is estimated to be below 10\%. 

Operational risks, e.g. failure of any active component (kicker, magnets, etc.) has no impact on the beam to SASE1 and SASE3. A vacuum failure in the transfer line can be quickly recovered by isolating the transfer line from the remaining vacuum system.  The 18~m new vacuum system in the XTL is of comparable complexity to the existing vacuum system and thus poses no additional risk. Only the two additional ceramic vacuum chambers for the kicker magnets add some complexity. European XFEL has already 18 of these chambers installed in different sections of the machine with no recorded issues so far.

Commissioning of the extraction line is considered to be straightforward. Experience with the commissioning of the already existing extraction lines and beam transports lead to a conservative estimate of a maximum of 6 dedicated commissioning shifts needed before the beamline can be operated in a complete parasitic mode.
\newpage
\section{Laser Specfications and Diagnostics}
\label{sec:laser}

\subsection{High Power Laser Requirements}
The main aim of the proposed experiments is to study the interaction of a high-intensity laser field with either an ultra-relativistic electron beam  or a high-energy photon beam. In particular, we aim at achieving conditions where the laser dimensionless amplitude $\xi$ significantly exceeds 1 and the electric field experienced in the electron rest frame or photon centre-of-momentum frame can be tuned from a small fraction of 1 up to a few. Due to the wide commercial availability and high performance at high power, we will consider from now on a Ti:Sapphire (Ti:Sa) system, operating at a central wavelength of $\lambda_L =$ 800 nm. Assuming a gaussian laser focal spot with FWHM of 8 $\mu$m, the condition $\xi>1$ (or, equivalently, $I_L > 4\times10^{18}$ Wcm$^{-2}$), translates to a laser power, in focus, of approximately 3 TW. Typical short-pulse Ti:Sa high-power lasers produce pulses as short as 17 fs. However, such a short pulse duration is prone to non-negligible shot-to-shot fluctuations and we will run at a more conservative duration of the order of 30 fs. A 3 TW pulse with a 30 fs duration will then contain, approximately, 100 mJ. It is a great experimental challenge to contain all the laser energy in a tight focus and typical high-power systems can only achieve 30 - 40 \% of the laser light within the FWHM of the laser focal spot, meaning that 100 mJ in focus needs a laser able to provide at least 300 mJ after compression (10 TW at 30 fs). Due to the non-perfect energy efficiency of laser compression systems (typically in the range of 60 - 70 \%), achieving $\xi>1$ in focus thus implies that the laser amplification chain needs to be able to provide at least 500 mJ before compression. Based on similar consideration, achieving $\xi\approx10$ would instead require a laser power, before being focused, of the order of $5$ J . Using 30~fs lasers with 0.9~J and 9~J, corresponding to 30~TW and 300~TW respectively, these requirements can be safely achieved. 

Assuming then $1<\xi<10$ and an electron energy of 14 GeV (Lorentz factor $\gamma_e = 2.74\times10^4$), translates into $0.16<\chi<1.6$, well within the requirements of the experiment. For electron energies of 17.5 GeV (Lorentz factor $\gamma_e = 3.42\times10^4$), the quantum parameter $\chi$ will instead be in the range $0.2<\chi<2$.

\begin{table}[ht]
\begin{center}

\begin{tabular}{|l|c|c|c|}
\hline
& \textbf{30 TW, 8$\mu$m}  &  \textbf{300 TW, 8$\mu$m}  &  \textbf{300 TW, 3$\mu$m}\\
\hline
& & & \\ 
  \textbf{Laser energy after compression (J)}  & 0.9  & 9 & 9 \\
  \hline
   \textbf{Percentage of laser in focus (\%)}  & 40  &  40 &  40 \\   
   \hline
    \textbf{Laser energy in focus (J)}  &  0.36 &  3.6  &  3.6 \\  
    \hline
    \textbf{Laser pulse duration (fs)}  &  30 &  30 &  30\\  
    \hline
    \textbf{Laser focal spot FWHM ($\mu$m) } & 8  &  8  &  3\\  
    \hline
    \textbf{Peak intensity in focus (Wcm$^{-2}$)}  &  $1.6\times10^{19}$ & $1.6\times10^{20}$ & $1.1\times10^{21}$ \\  
    \hline
    \textbf{Dimensionless peak intensity, $\xi$}  &  2 &  6.2 &  16\\   
    \hline
    \textbf{Laser repetition rate (Hz)}  & 1  &  1  &  1\\  
    \hline
     \textbf{Electron-laser crossing angle (rad))}  &  0.35 &  0.35 &  0.35\\  
     \hline
     \hline
      \multicolumn{4}{c}{}   \\
    \multicolumn{4}{l}{\fbox{\textbf{14 GeV electrons}}}   \\
  
    \multicolumn{4}{c}{}   \\
      \hline
     \textbf{Electron Lorentz factor}  &  $2.7\times10^4$ &  $2.7\times10^4$ &  $2.7\times10^4$ \\ 
 \hline
    \textbf{Quantum parameter $\chi$}  &  0.32 &  1 & 2.6\\  
    \hline
    \hline
    \multicolumn{4}{c}{}   \\
    \multicolumn{4}{l}{\fbox{\textbf{17.5 GeV electrons}}}   \\
   
    \multicolumn{4}{c}{}   \\
     \hline
     \textbf{Electron Lorentz factor}  & $3.4\times10^4 $ & $3.4\times10^4 $  & $3.4\times10^4 $\\ 
 \hline
    \textbf{Quantum parameter $\chi$}  &  0.41 &  1.26 &  3.26\\  
    \hline
    \hline
   
\end{tabular}
\caption{Typical laser and interaction parameters for different initial laser power and focussing geometry, for two different electron energies: 14 GeV and 17.5 GeV.}\label{table1}
\end{center}
\end{table}

We then propose a stepped approach where, in the preparatory phase, a 30 TW Ti:Sa system focussed down to a focal spot with FWHM of 8 $\mu$m will be used (first column in table \ref{table1}). This will imply, for a 14 GeV (17.5 GeV) electron beam, $\xi = 2$ and $\chi = 0.3$ ($\chi = 0.4$). Such a smaller laser system will be sufficient for the first stage of the experiment, allowing one to perform precision studies of non-linear Compton scattering, quantum radiation reaction, and to perform diagnostic tests and background optimisation as well as initial studies of non-linear Breit-Wheeler pair production. The key parameters are in the vicinity of those of the E144 experiment. 

In a second phase, the laser power will be upgraded to 300 TW, which will imply, for a 14 GeV (17.5 GeV) electron beam, $\xi = 6.2$ and $\chi = 1$ ($\chi = 1.3$). As a final stage we will then reduce the focal spot size, in order to increase the peak intensity in focus. Achieving a focal spot FWHM of 3 $\mu$m will then result, for a 14 GeV (17.5 GeV) electron beam, to $\xi = 16$ and $\chi = 2.6$ ($\chi = 3.3$). These considerations are summarised in table \ref{table1}, for an electron beam energy of 14 GeV or 17.5 GeV.

\subsection{High Power Laser Technology}
\begin{figure}[ht]
    \centering
    \includegraphics[width=0.7\textwidth]{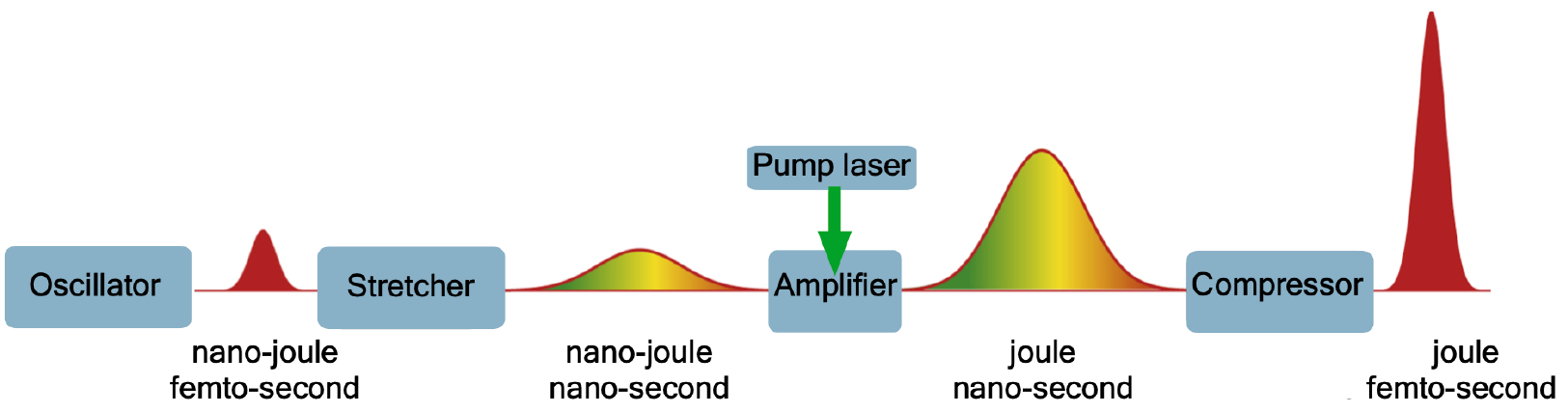}
    \caption{Cartoon depiction of the chirped pulse amplification (CPA) technique.}
    \label{fig:cpa-cartoon}
\end{figure}

The high-power laser system (HPLS) will utilize the chirped pulse amplification (CPA) technique \cite{strickland1985compression}, which is illustrated in Fig. \ref{fig:cpa-cartoon}.  An ultrashort, low-energy laser pulse is stretched in time, amplified up to the required energy, and then compressed back. This way, light level is kept below the amplifying media's damage threshold.  The result is high-peak power, fs-scale duration laser pulses. 

\begin{figure}
    \centering
    \includegraphics[width=0.7\textwidth]{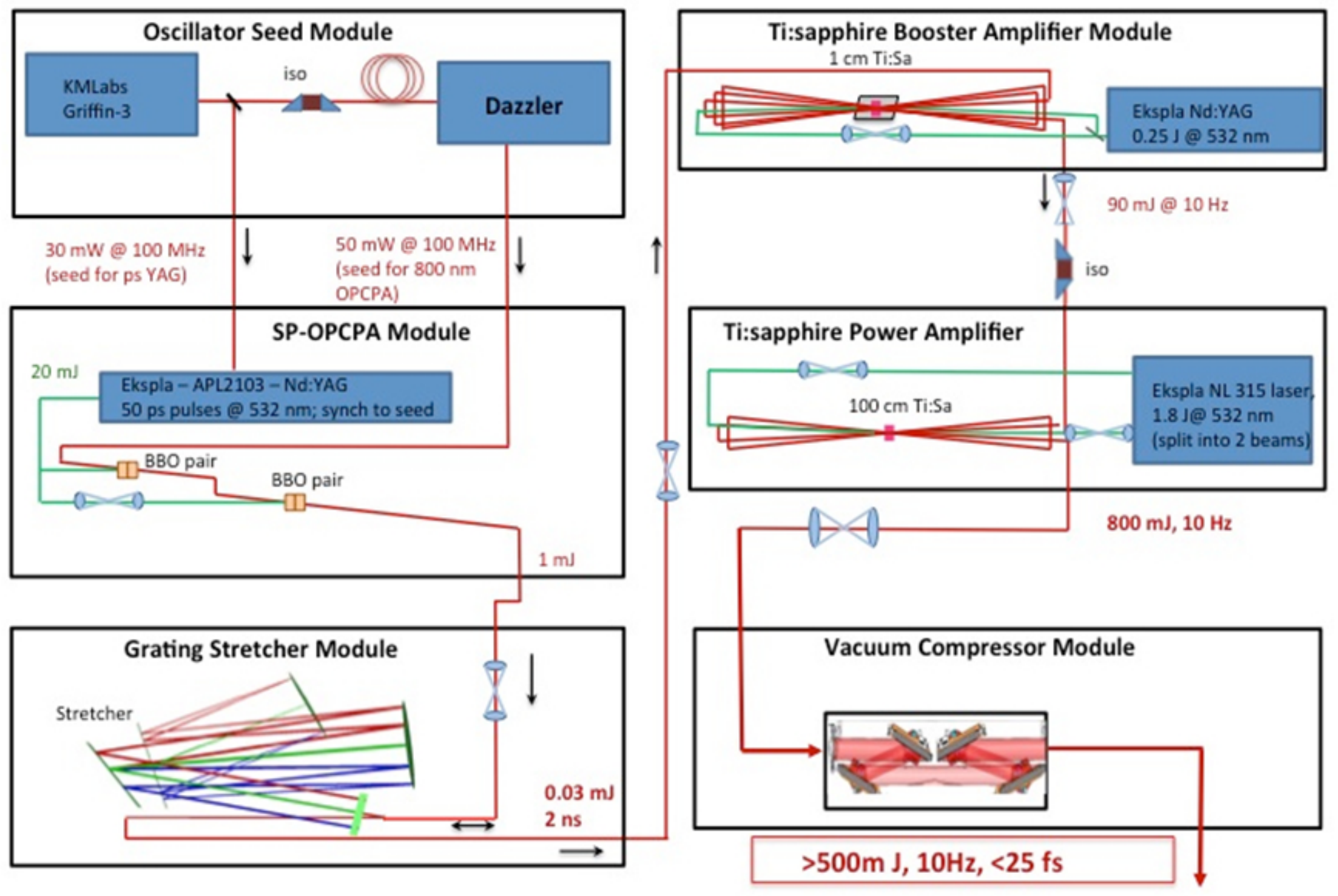}
    \caption{A typical optical layout of a high contrast HPLS.}
    \label{fig:optical-layout}
\end{figure}

A  typical optical layout of a commercially available high-contrast HPLS is shown in Fig. \ref{fig:optical-layout}, which includes the following components:

\begin{itemize} 
\item \textbf{Front end}\newline
The laser chain starts with a commercial femtosecond oscillator which delivers a $\approx 75$ MHz pulse train. Each of these pulses is a few-fs long with a few-nJ in energy. The system picks one of these pulses at a rate of $5 - 10$ Hz, and amplifies it to about 1 mJ. This part of the system is known as the front-end. Different amplification technologies may be employed here. They differ in their cost, reliability, and achievable pulse contrast; i.e. the light intensity level which precedes the main laser pulse. Our front-end will be based on either non-linear pulse cleaning \cite{jullien200510}, or on short-pulse optical parametric chirped pulse amplification \cite{musgrave2010picosecond}, both of which have proven to provide excellent contrast. A low contrast system can result in significant fraction of the pulse energy being outside the main pulse temporally, which is undesirable in the context of precision experiments.

\item \textbf{Optical pulse stretcher}\newline
Following the front-end, the pulses are sent into a grating-based pulse stretcher.  There they are steered to hit an all-reflective ~1500 lines/mm grating four times.  The stretcher bandpass is set to about 100 nm to avoid clipping effects that would reduce pulse contrast. 
\item \textbf{Multi-pass 10 Hz Ti:Sapphire power amplifier}\newline  
Following the stretcher, the pulses have approximately 0.3 mJ of energy.  These pulses are spatially filtered and amplified in 3–5 amplification stages. Each stage consists of a Ti:Sapphire crystal pumped with ns-long green (532 nm) laser pulses. Following each stage, the beam is expanded to remain below the damage threshold of the following optics in the optical chain. The number of amplifiers required in this section will depend on the final peak power to be achieved. For a 30 TW laser, three amplifiers will be typically needed, with this number increasing to 4 or 5 for the 300 TW system.
\item \textbf{Optical pulse compressor}\newline    
The fully amplified pulses will be expanded and sent through a window into an optical pulse compressor which operates under vacuum.  The compressor design is based on two large gold-coated diffraction gratings. The compressor design will also set the practical repetition rate.  Thermal aberrations limit the practical repetition rate to 1 Hz in the absence of grating cooling or active compensation. This ultimately limits the repetition rate of the whole system to about 1 Hz. 

\end{itemize}

In order to perform precision studies, it is crucial to have high quality diagnostics to closely monitor the effects of shot-to-shot fluctuations and long-term drifts on the precision of the data. High-power, femtosecond laser systems are precision tools and respond with significant performance changes to relatively small drifts in alignment. This is because small variations of spatial and spectral phase have noticeable effects on the peak intensity. Such variation can be caused by thermal effects, air currents and mechanical vibrations and drift.
Current commercial systems are engineered to consist of distinct mechanical components set out on a conventional optical table and must therefore be re-aligned routinely and are open to variety of effects varying performance. This contrasts with sealed, turnkey systems which exist in different parameter ranges in industry.
However, we believe that this will not limit the ultimate precision and reproducibility we can achieve. In the following we will discuss the considerations setting out the laser requirements and interaction geometry.

\subsection{Laser Diagnostics}

Typical shot-to-shot fluctuations are significant and require an approach to intensity tagging for each interaction shot. In the following we detail the typical level of shot-to-shot fluctuations and approaches with which we aim to tag each shot with a precision of below 0.1\%.
High energy Ti:Sapphire lasers using flash-lamp technology for the pump-lasers have typical energy fluctuations of around 2-3\% rms. These overall energy fluctuations, while significant, are not the dominant contribution to shot-to-shot intensity fluctuations. The major contributions to such fluctuations are small changes in phase both in real space and in frequency space. Small changes in spatial phase result in the spot radius fluctuating at the few \% level resulting in 10\% level intensity fluctuations. Similarly, fluctuations in the spectral phase can lead to 5-10\% rms fluctuations in the pulse duration. In total, the shot-to-shot variation in intensity $I_0$ on a stable laser can reach 15\% or more. 
To mitigate against this, we will set-up a state of the art diagnostic system capable of measuring the fluctuations. The shots will then be tagged with their precise relative intensity allowing precision relative measurements using the following diagnostics:
\begin{itemize} 
\item \textbf{Energy Tagging}\newline
The full beam energy will be measured by imaging an attenuated beam onto a CCD. For a well exposed image containing $>10^{10}$ total counts the energy fluctuations can be controlled to $<10^{-5}$ accuracy.

\item \textbf{Fluence Tagging}\newline
A similar approach will be taken to determine the fluence $F_{\omega}=\frac{dE}{dA}$ using a high magnification image of the focal spot onto a CCD camera. Care will be taken to eliminate any non-linear effects in transmissive optics. The variation of the efficiency of the CCD with wavelength would reduce the effectiveness of the method described above if there are significant shifts or fluctuations of the spectrum. We will employ a spectrometer and colour filtered images to maintain the high precision that is theoretically possible in these measurements.

\item \textbf{Pulse Length Tagging}\newline
Finally, the stability of the pulse duration will be determined by employing two complementary techniques. We will measure the pulse duration on every shot using a state of the art system capable of reconstructing the full pulse shape such as the Wizzler combined with an autocorrelator. We will employ a simpler, complementary technique to ensure that this measurement is not dominated by fluctuations within the measurement device by producing an image of the frequency converted beam and comparing it to the corresponding image of the fundamental beam. Since the conversion in a thin non-linear crystal far from saturation scales as $I^2$ any fluctuations will appear directly as variations in the ratio of the two images $R_{\omega,2\omega}=F_\omega/F_{2\omega}$. Calibration of this diagnostic should allow precise relative pulse duration tagging. Although we do not anticipate this to achieve the level of precision of the energy and flux diagnostics, a relative pulse duration precision of $<10^{-3}$ is deemed achievable. Finally, we will measure fluctuations in the angular chirp using imaging spectrometers.

\end{itemize}

\begin{figure}[ht]
    \centering
    \includegraphics[width=0.4\textwidth, angle=90]{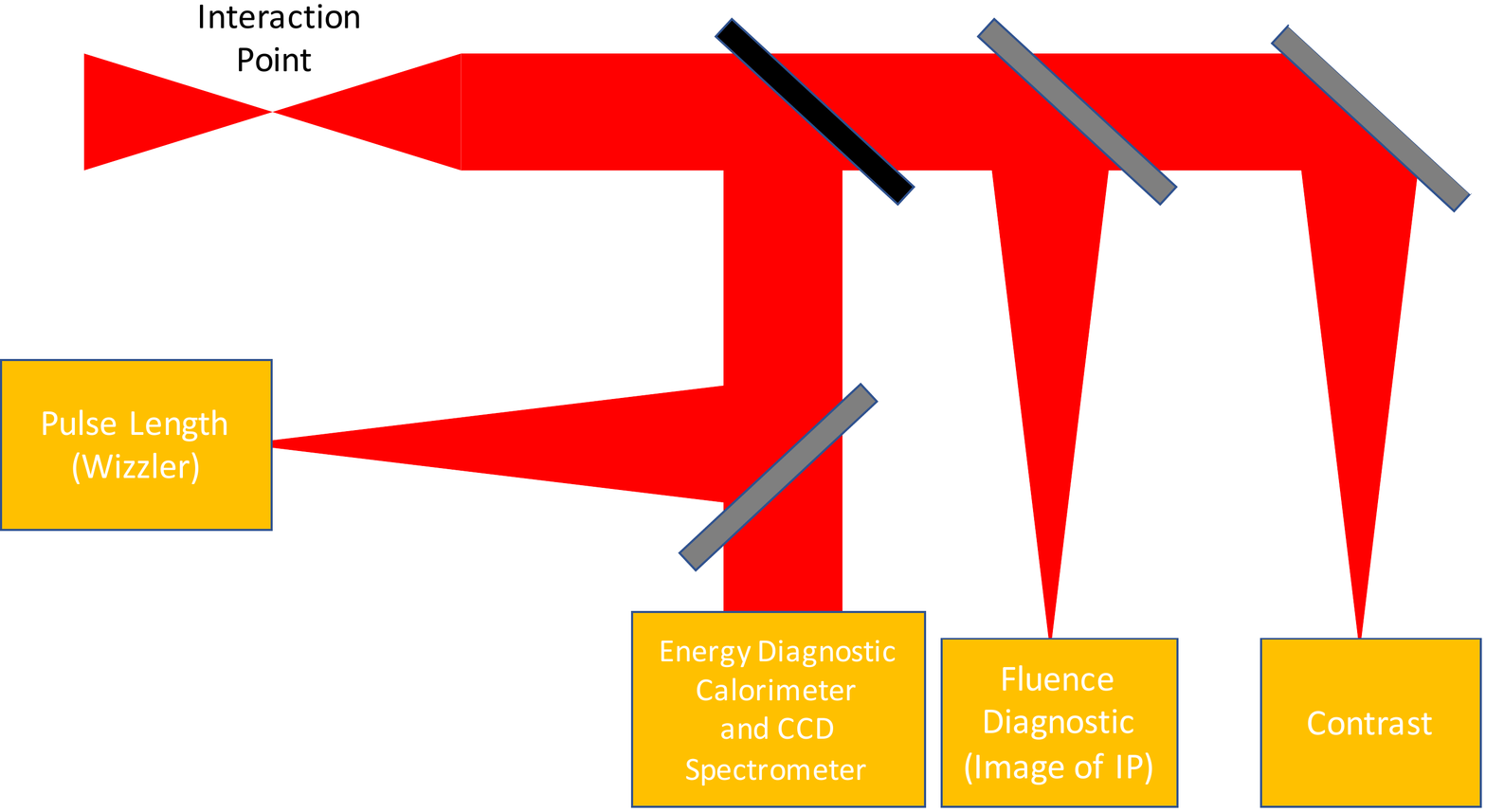}
    \caption{Schematic of the proposed intensity tagging diagnostics. The laser beam will be transported and attenuated on the path to the diagnostics.}
    \label{fig:diag_layout}
\end{figure}

\subsection{Determination of the Peak Electric Field in Focus}

\begin{figure}[ht]
    \centering
    \includegraphics[width=1\textwidth, angle=0]{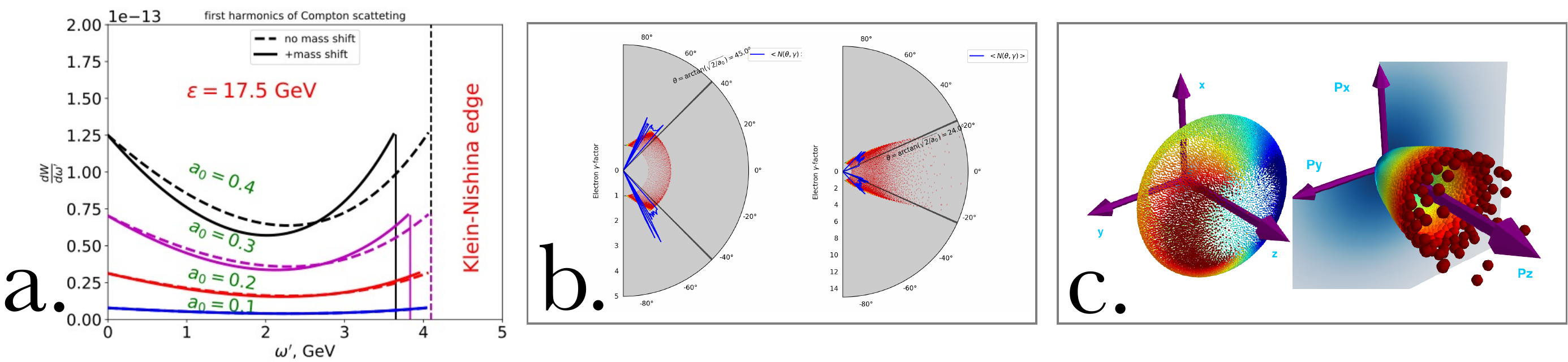}
    \caption{ Examples of different indirect methods to measure the peak intensity of the laser in focus: \textbf{a.}~Simulated spectra of scattered photons for differing values of $\xi$ (referred to as $a_0$ in the labelling). The edge shift is visible even at $\xi<1$ and can in principle be resolved in the downshifted electron spectra with high accuracy. \textbf{b.}~Angular distribution of electrons initially at rest for $\xi = 2, 10$. The blue line shows the energy of the averaged number of electrons scattered under a given angle. A well defined edge depending on $\xi$ is visible allowing the effective peak field strength to be cross-calibrated in the focus of the laser. \textbf{c.}~Angular distribution of electrons scattered by a laser pulse in real and momentum space. In this example, the laser dimensionless parameter is $\xi = 10$ and the focal spot has a FWHM of $4 \, \mu m$. }
    \label{fig:xi}
\end{figure}

In order to provide precise information on the interaction and therefore perform rigorous comparison with available strong field QED models, it is necessary to accurately measure the laser peak intensity in focus. 
In principle the tagging system outlined above will provide an accurate measurement of the peak field in the interaction point. We aim to cross-calibrate our diagnostic suite described above at the Light Intensity Unit of the PTB Braunschweig~\cite{ptbbraunschweig}, which can provide traceable standards for CW sources to SI Units with $10^{-4}$ accuracy. This will provide tight limits on the actual value of $\xi$.  However, a systematic offset due to small imaging aberrations of the focal intensity distribution at the interaction point and spectral phase errors in the transport will be unavoidable and is not easily accounted for. It is thus desirable to have a direct measurement of the peak intensity in focus and different methods have been proposed in the literature. For instance, it has been recently shown \cite{Quere:NPhot} that spatially resolved Fourier-transform spectroscopy allows one to reconstruct the spatio-temporal profile of the laser electric field in focus. This method can be compared with other indirect methods (see Fig. \ref{fig:xi}), in order to provide absolute characterisation of the laser beam in focus. 
For instance, one might exploit the well-known strong correlation between the Compton edge shift and $\xi$ (Fig. \ref{fig:xi}.a). Alternatively, we can measure the scattering of electron initially at rest from a tenuous Helium target at the IP (for these measurements the IP will be vacuum isolated from European XFEL). The electrons are scattered at well defined angles dependent on the laser intensity (Fig. \ref{fig:xi}.b,c). These scattering angles are outside the beam cone and can be detected without modifying the optical layout.  Based on our  simulations of Compton shift and ponderomotive scattering we aim to achieve an absolute calibration of better than 5\% in the early phase of LUXE with continual improvements towards 0.1\% absolute accuracy in the laser E-field appearing realistic.

\subsection{Timing of Laser versus Electron Beam}
The performance of the experiments requires the overlap in space and in time at the IP of the
laser pulse with the LINAC electrons (or the Bremsstrahlung gammas). Overlap in space
requires an alignment, positioning and drift compensation system for the laser focus, to track
the electron bunches, with transverse dimensions of order 20~$\mu$m. The timing system must
ensure simultaneous arrival at the IP of the laser photon pulse (with nominal 30 fs
duration) and the electron bunch (with typical length of $\sim 40$~$\mu$m, corresponding to $\sim 130$~fs
duration). Considering the geometry of the electron – laser collisions at angle about $theta
\simeq 0.3$~rad from the head-on condition, a timing accuracy of a few 100s fs is required.
Analogous, although less stringent (due to the longer laser and electron pulses), 
requirements were faced for the E144 experiment at SLAC. At the European XFEL the
extreme requirements of optical pump – $X$-ray probe experiments (down to $20$~fs) led to the
development of an extensive synchronisation system~\cite{Synchroxfel}, based on an ultra-stable, 
low-noise laser oscillator, whose signals are distributed via actively length-stabilised
optical fibres to the different locations across the accelerator and experimental areas.
There, it is used to locally re-synchronise radio frequency signals, to precisely measure the
arrival time of the electron beam for fast beam-based feedbacks, and to phase-lock
optical laser systems for electron bunch generation, beam diagnostics and user pump-probe
experiments. It is foreseen to make use of this existing system to synchronise the optical
laser to the electron bunch arrival time.
High-power laser systems typically suffer from electronic shot-to-shot temporal jitters and,
potentially, slow temporal drifts that must be taken into account for this class of experiments.
While a precise control of temporal drifts involving active feedbacks can ensure sufficient
stability levels -- as recently implemented and successfully tested at the Central Laser Facility
in the UK -- temporal jitters cannot be completely ruled out. Baseline commercially available
high-power laser systems can now guarantee jitters at a sub-ps level with tens to hundreds of
femtoseconds potentially achievable. This level of temporal stability is suitable for the aims
set by the LUXE experiment. As far as the jitter in the arrival time of electron bunches is
concerned, active feedbacks reduce the jitter to order 13~fs over a few hundred bunch trains, and even in the absence of feedbacks, drifts are limited to 450~fs peak to peak variations over a period of two days~\cite{Schulz}.

\subsection{Definition of Requirements}
The laser performance required takes into account the following real-world estimates for inefficiencies and transport losses: we estimate that the peak achievable intensity is no more than 40\% of the theoretical value (see above). Part of the beam energy is lost from the interaction point through beamline transport losses and Strehl ratio $<$1 (i.e. spatio-temporal phase errors).

A 30 fs laser beam with $E\approx 1$~J is now readily available commercially and has a footprint of the order of 20~m$^2$. Upgrading to a 300~TW system will increase the footprint to approximately 40~m$^2$. The main requirements beyond the modest power and cooling water of such a laser are the availability of a temperature and humidity controlled  clean-room environment to house the laser with a separate enclosure for pumps and power supplies.

We note that even though the first proof of concept experiments are possible with smaller systems (30 TW), it should be in principle set up already for a swift upgrade to higher power. However, some additional civil construction may be required to accommodate the larger system.   

\newpage
\section{Detectors, Monitors and Data Acquisition Aspects}
\label{sec:detectors}
Three detector subsystems are needed to measure the $\gamma_B + n\gamma_L$ process:  the region after photon production via Bremsstrahlung; electron--positron pair production after the interaction point (IP); and photons which continue downstream to the forward photon spectrometer.  In the case of the $e + n\gamma_L$ process, only the final two detector subsystems are required, although the electron detector system after the IP will need to be modified as the rate of electrons will be high (see Table~\ref{tab:egammarates}).  The final-state particles that need to be measured are electrons, positrons or photons, with the major challenge arising from the potentially very large number of final-state electrons and photons and the backgrounds from secondary interactions.  The detectors must also have a high granularity so that the number of particles can be counted reliably, even at high intensity, and must have a linear response over a wide dynamic range.

In the following sections, the particle rates are discussed as well as the three regions or detector subsystems on a global level and initial designs are presented.  The detector technology solutions which can be common to more than one subsystem are then presented.  Finally the data acquisition system is briefly discussed.

\subsection{Particle Rates}

Tables~\ref{tab:egammarates} and \ref{tab:gammagammarates} show the estimated rates of electrons, positrons and photons in the various detector regions for the $e+n\gamma_L$ and $\gamma_B+n\gamma_L$ setup, respectively. It is seen that they span a wide range from values $\ll 1$ up to $\sim 10^9$, placing very difficult requirements on the technologies for detection of particles and the measurement of their energies.

\begin{table}[htbp]
    \centering
    \begin{tabular}{l|c|c|c}
         Location & particle type & rate for $\xi=2.6$ & rate for $\xi=0.26$   \\\hline\hline
         $e^-$ detector & $e^-$, $E_e<16$~GeV & $1.5\times 10^9$ & $6\times 10^6$ \\
         $e^+$ detector & $e^+$ & $15.3$ & $<0.01$ \\
         Photon detector & $\gamma$ & $6\times 10^{10}$ & $1\times 10^7$ \\
         Photon detector (W foil) & $e^+$ and $e^-$ & $6\times 10^6$ & $1\times 10^4$  \\
         Photon detector (W wire) & $e^+$ and $e^-$ & $1.5\times 10^5$  & $1\times 10^2$  \\
         \hline
         \\
    \end{tabular}
    \caption{Rates of particles for the $e$--laser setup. The two signal processes are the Compton and the Trident process. Also shown are background rates. In all cases $E_\textrm{beam}=17.5$~GeV and $1.5\times 10^9$ electrons/bunch are assumed.
    \label{tab:egammarates}
    }
\end{table}

\begin{table}[htbp]
    \centering
    \begin{tabular}{l|c|c|c|}
         Location & particle type & rate for $\xi=6.5$ & rate for $\xi=1.2$  \\\hline\hline
         $e^-$ detector behind converter& $e^-$, $E_e<13$~GeV & \multicolumn{2}{c|}{$2\times 10^7$} \\
         $e^+$ detector behind converter& $e^+$ & \multicolumn{2}{c|}{$9\times 10^4$} \\
         photons after converter & $\gamma$ & \multicolumn{2}{c|}{$1.3\times 10^8$} \\
         $e^\pm$ detector behind IP& $e^-/e^+$ & $350$ & $1\times 10^{-2}$ \\
         Photon detector & $\gamma$ & \multicolumn{2}{c|}{$1.3\times 10^8$} \\
         Photon detector & $e^+$ and $e^-$ & \multicolumn{2}{c|}{$160$} \\
         \hline
         \\
    \end{tabular}
    \caption{Rates of particles for the $\gamma_B$--laser setup. In all cases $E_\textrm{beam}=17.5$~GeV and $1.5\times 10^9$ electrons/bunch are assumed.
    All rates are before any geometric acceptance cuts.
    \label{tab:gammagammarates}
    }
\end{table}

\subsection{Detector Regions}
To provide precise counting and energy measurement of the $e^+e^-$ pairs from collisions at the interaction point, a few layers of silicon pixel tracking detector followed by a few layers of calorimeter detector, are placed downstream the beamline on its two sides. This system is adequate for the relatively low rates of particles (see tables~\ref{tab:egammarates} and ~\ref{tab:gammagammarates}) of typically $<100$. Not only can the proposed combination of a tracker and a calorimeter provide a precise measurement of the produced $e^+e^-$ pairs, it can also enable a strong background suppression via coincidence and matching requirements, in and between the two subsystems. The two subsystems can also be used to cross-calibrate each other such as the energy scales and resolution. The calorimeters will also absorb all particles and so act as an effective dump.

For the detection of $e^+e^-$ pairs behind the tungsten converter, the particle rates are $10^4-10^6$ and Cherenkov detectors will be used. These are more adequate for such high particle fluxes and more cost effective. The purpose of this system is to provide an \textit{in-situ} measurement of the photon flux. Cherenkov detectors will also be deployed for the electron detector behind the IP in the $e$--laser setup. 

The dimensions of the sensitive elements of the detector systems are about 50~cm in the transverse plane and at least $\sim 1.5$~cm in the vertical plane.

The detector readout is to be triggered either by the laser pulse or by the beam traversing the IP.
The main physics measurement is performed when both the laser and the beam are present while backgrounds can be measured if only one is present and the other is absent. The maximum rate expected is 10 Hz.  With each such trigger, the detector is read out and the raw information is stored for later analysis.

In principle, a given subsystem could be read out using one data acquisition computer.

\subsubsection{Photon Production via Bremsstrahlung}
\begin{figure}[htbp]
  \begin{minipage}[t]{0.47\textwidth}
    \includegraphics[width=\textwidth]{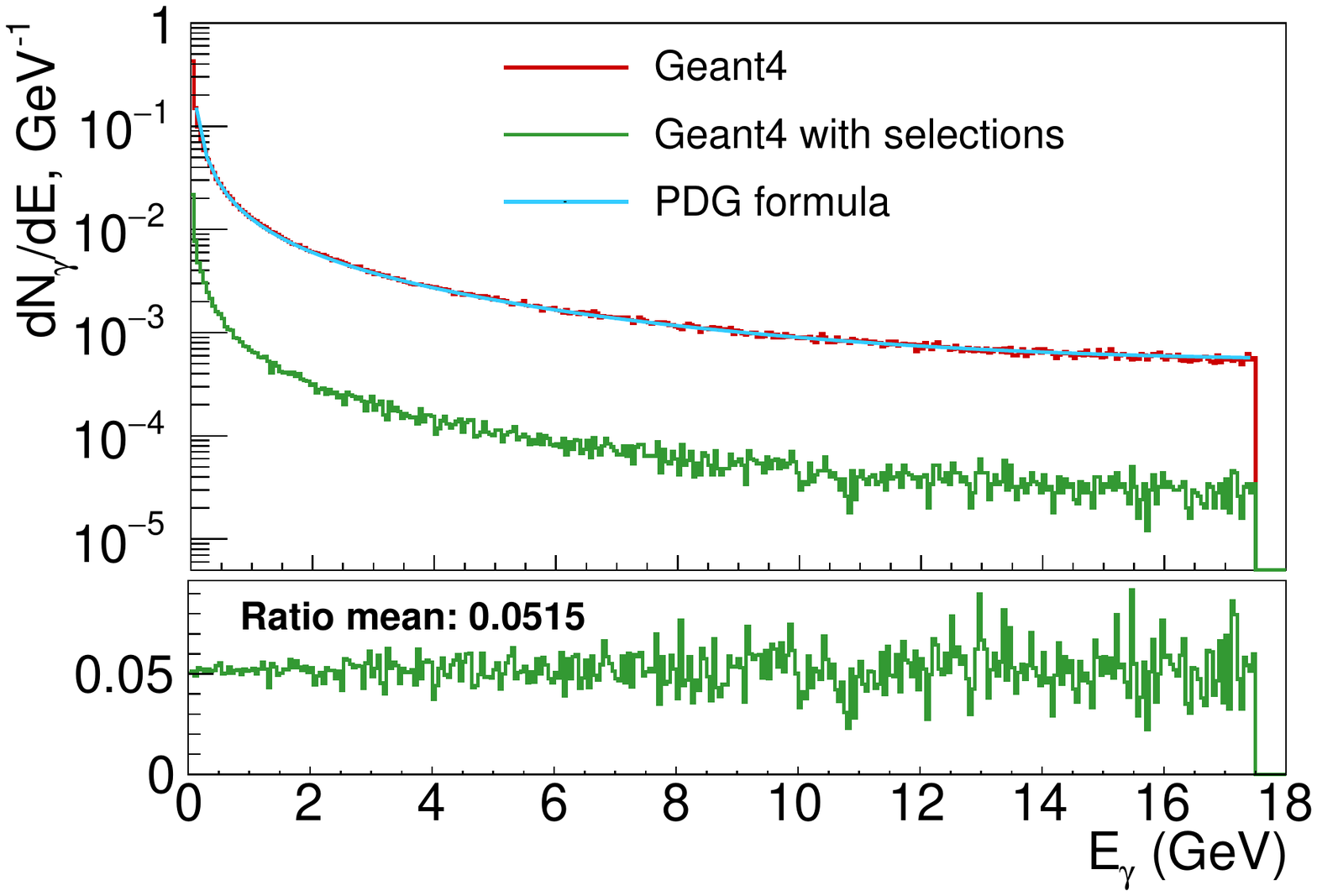}
    \caption{Comparison of Bremsstrahlung spectra obtained from theoretical formula in the PDG~\cite{Tanabashi:2018oca} and with GEANT4~\cite{Allison:2016lfl,Allison:2006ve,Agostinelli:2002hh} simulations for a 
    tungsten target of 35\,$\mu$m (1\%$X_{0}$) thickness. The green line shows the $\gamma$ spectrum after imposing limits on position in the 
    transverse plane to $\pm$25\,$\mu$m. 
    The bottom plot shows the fraction within $\pm$25\,$\mu$m compared to the total.
    \label{fig_brems_spectra_g4_vs_pdg}}
  \end{minipage}\hfill
  \hspace{0.05\textwidth}
  \begin{minipage}[t]{0.47\textwidth}
    \includegraphics[width=0.99\columnwidth]{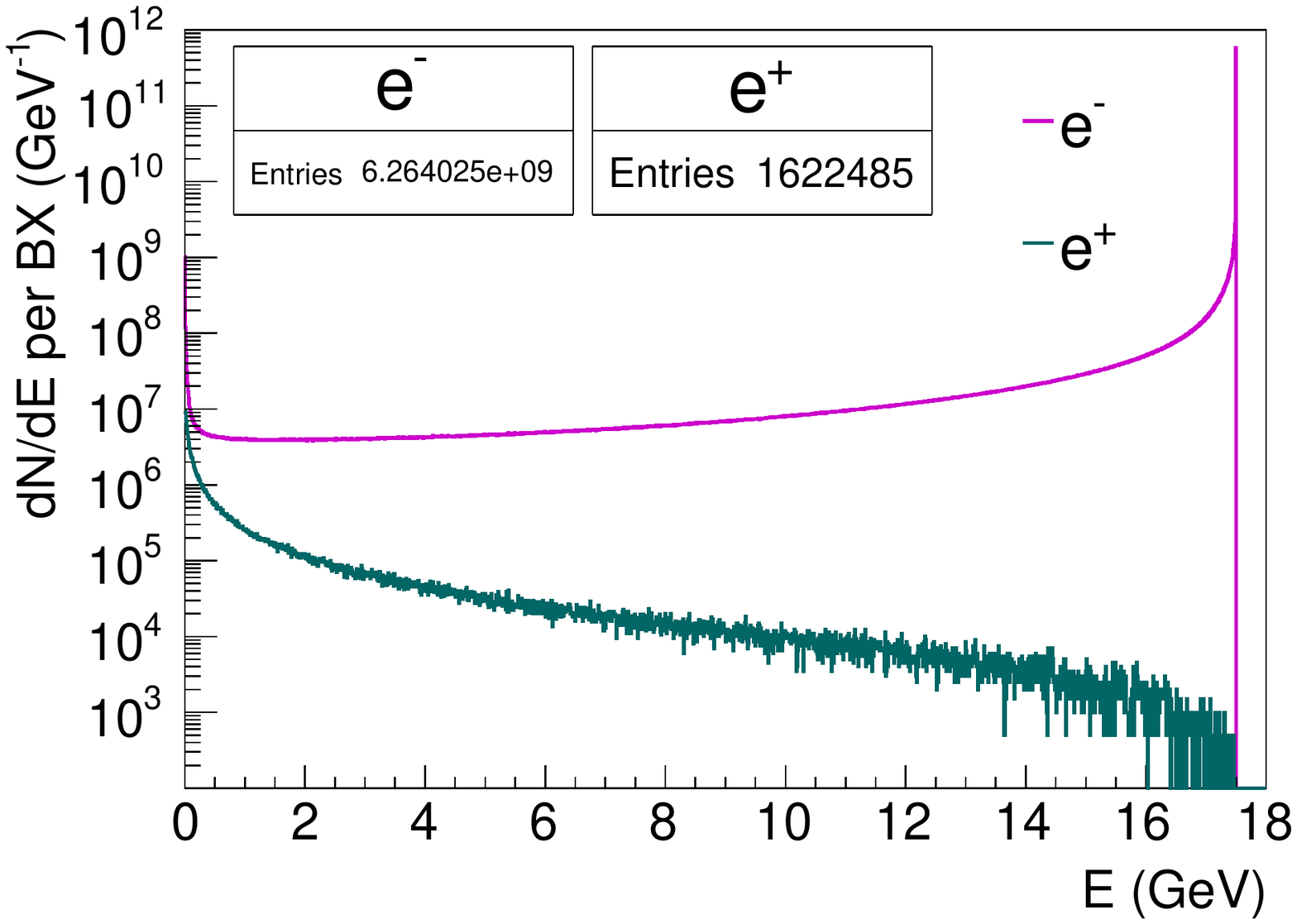}
    \caption{Energy spectra of electrons and positrons observed in the simulation with a tungsten target of 35\,$\mu$m (1\%$X_{0}$) thickness for one 
    bunch crossing. The electron spectrum also includes electrons from the beam which did not interact with the converter.}
    \label{fig_el_pos_spectra}
  \end{minipage}\hfill
\end{figure}

In order to study the process of electron--positron pair production via a photon absorbing multiple low energy photons, i.e. 
$\gamma_B +n\gamma_L \to e^+e^-$, a beam of high energy photons is required.  This is achieved using a photon converter which is placed in the 
path of the extracted European XFEL electron bunch yielding a broad-band spectrum of photons via Bremsstrahlung.  The converter is a 35\,$\mu$m thick 
tungsten target; the photon energy spectrum is shown in Fig.~\ref{fig_brems_spectra_g4_vs_pdg}.  The expected spectrum from a GEANT4 simulation and the empirical prediction~\cite{Tanabashi:2018oca} agree well.  The final fraction of these photons that interacts with the laser beam, i.e.\ with a position in the transverse plane limited to $\pm 25\,\mu$m is about 5\%.

\begin{figure}[htp]
   \includegraphics[width=\textwidth]{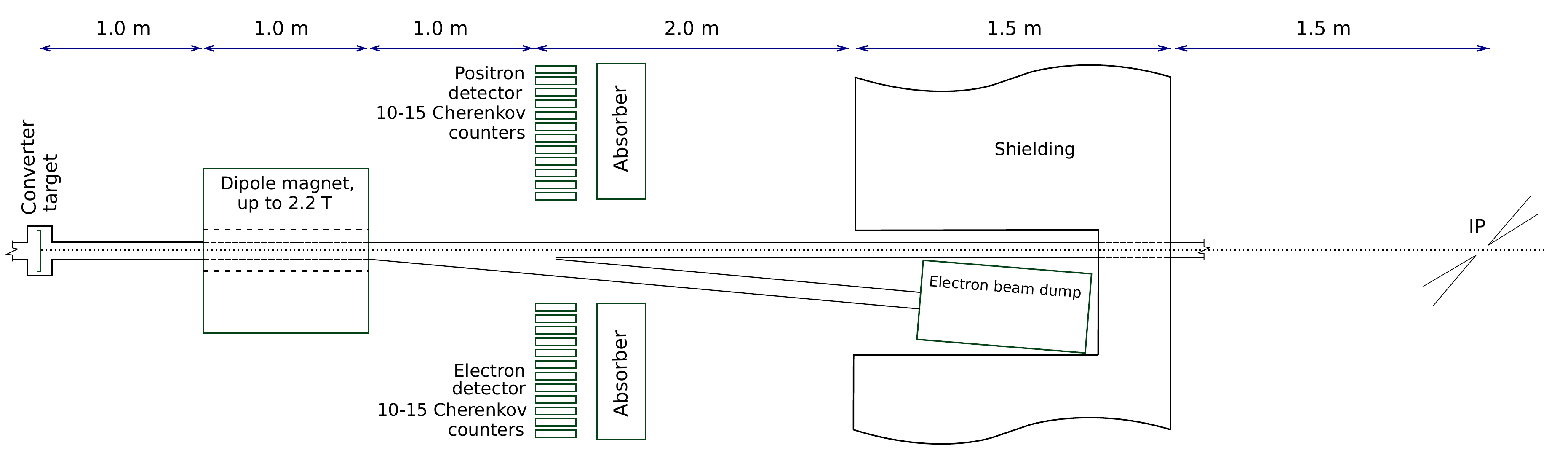}
   \caption{Schematic of the area around the photon target during $\gamma_B$--laser running.  After the target, a high-field dipole magnet is placed 
   to separate electron--positron pairs produced in the target as well as the electrons from the initial beam which underwent Bremsstrahlung to some degree.
   Cherenkov counters are foreseen to measure electrons and positrons with absorbers placed behind these to measure the total energy and serve as a dump for the electrons and positrons.  An electron beam 
   dump to capture the high energy electrons follows, as well as shielding to protect the IP from stray radiation. 
   }
   \label{fig:brems}
\end{figure}

The expected average number of Bremsstrahlung photons produced per bunch crossing (BX), assuming $1.5\times 10^9$ electrons per bunch,  
is about $1.5 \times 10^8$, the number of positrons is about 
$1 \times 10^5$ and the total number of electrons with $E<13$~GeV observed behind the target is $2 \times 10^7$, dominated by electrons from the beam which have lost some energy due to Bremsstrahlung. The positrons are coming from pair production of the Bremsstrahlung photons within the tungsten foil. 
The spectra of positrons and electrons and 
their average numbers for one bunch crossing are presented in Fig.~\ref{fig_el_pos_spectra}.  Detectors to measure the number of electrons and positrons 
after the tungsten target are required to act as monitors for the number of high energy photons produced.
It is envisaged to use Cherenkov detectors for both the electrons and the positrons as the fluxes are high. Since they are behind a magnet the location of the electron is a measure of the energy. The granularity of the Cherenkov counters is chosen as $2 \times 2$~cm$^2$ and about 15 units are anticipated. Such a setup is shown schematically in Fig.~\ref{fig:brems} where a high-field dipole magnet is used to separate electrons and positrons. A beam dump and shielding protect the IP from unwanted stray radiation.

\begin{figure}[htp]
   \includegraphics[width=0.45\textwidth]{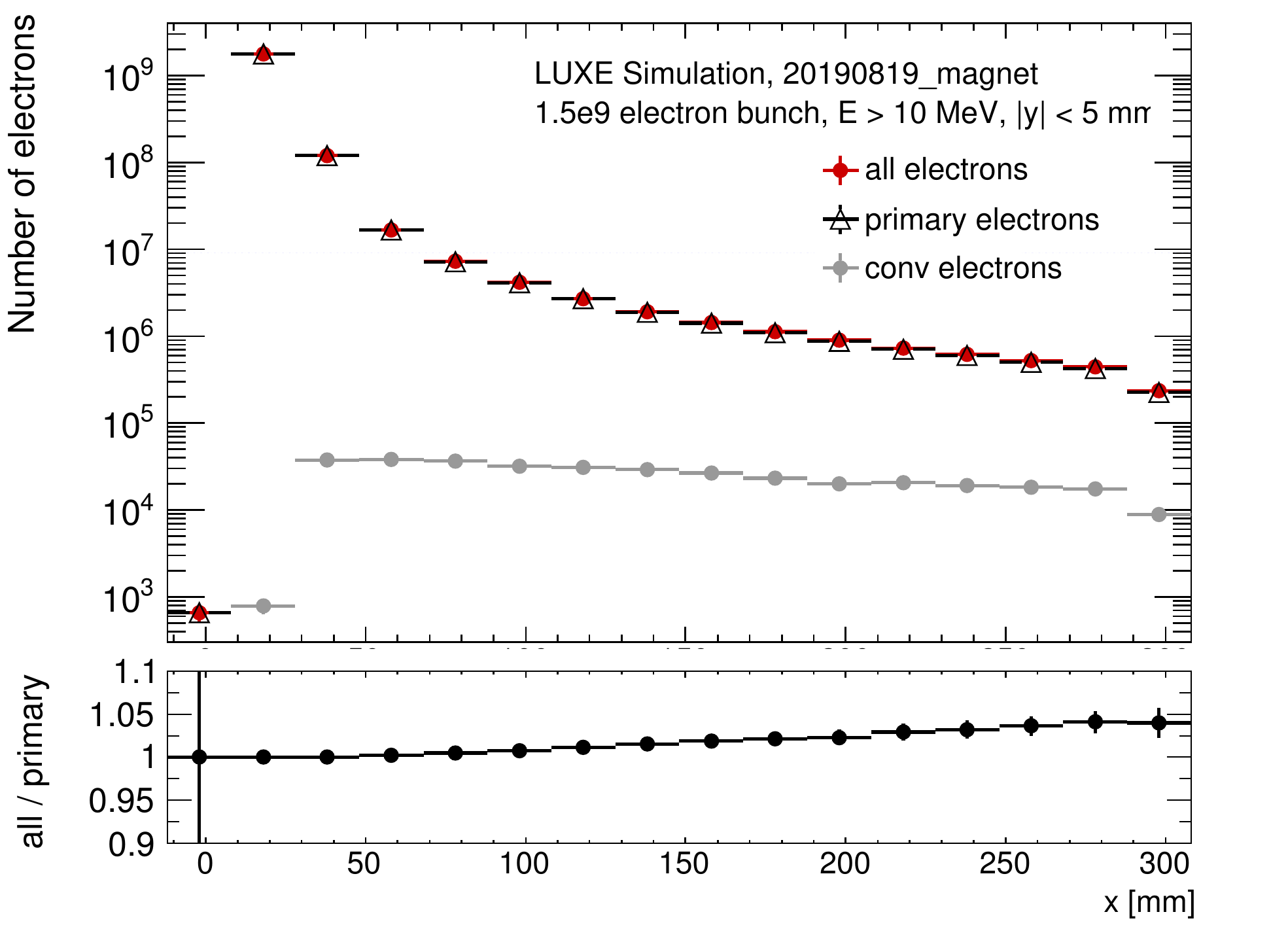}
   \includegraphics[width=0.45\textwidth]{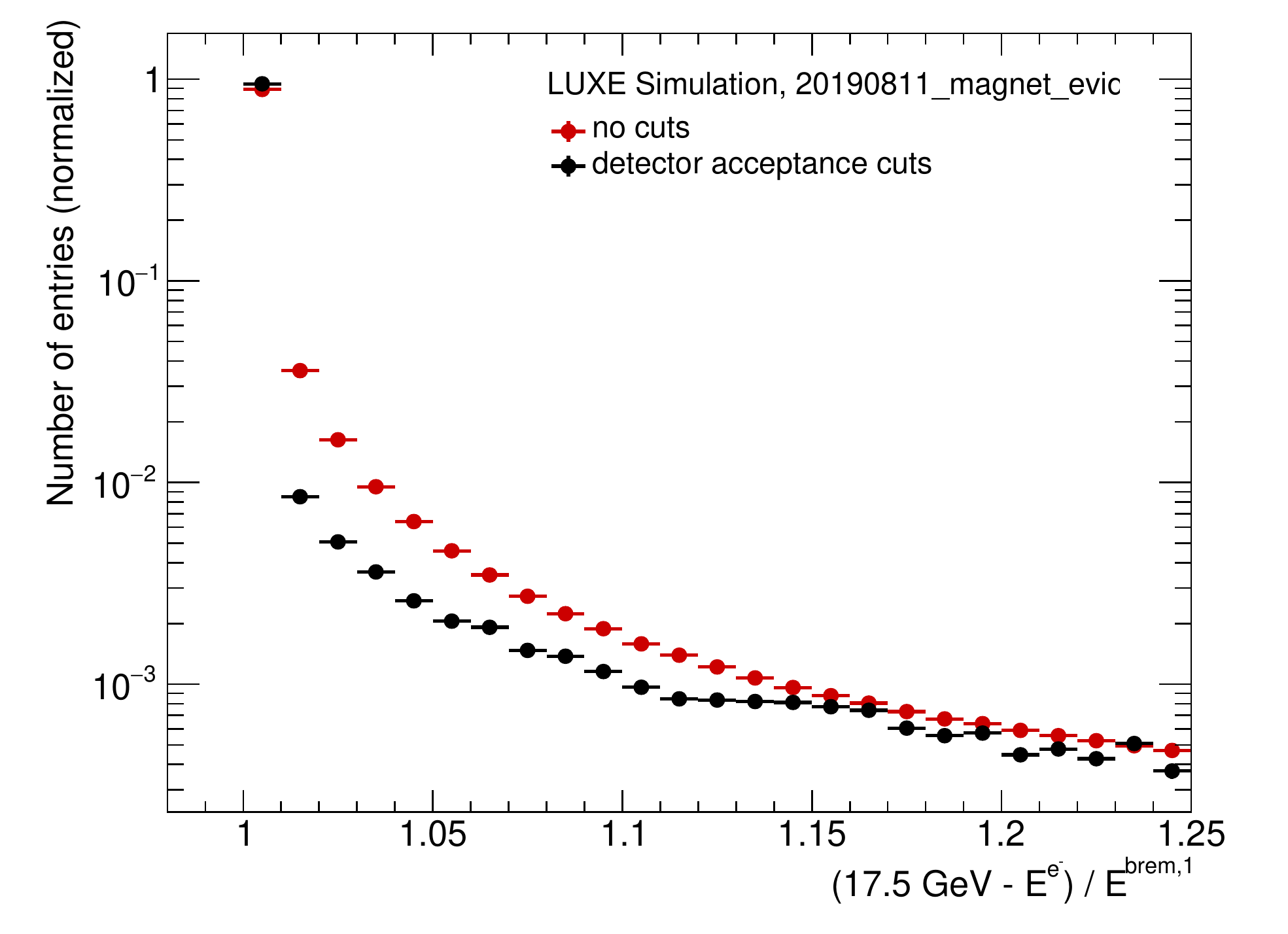}
   \caption{Left: Number of electrons versus $x$.
   Right: Ratio of the reconstructed photon energy to the true photon energy: $(E_\textrm{beam}-E_e)/E_\gamma$. Less than 2\% of the events have values $(E_\textrm{beam}-E_e)/E_\gamma>1.25$. Here, $E_\gamma$ is the true energy of the highest energy photon in the event.
   \label{fig:bremrec}
   }
\end{figure}

The measurement of the electron energy spectrum can be used to determine the $\gamma_B$ energy spectrum as the photon energy is roughly given by $E_{\gamma_B}=E_\textrm{beam}-E_{e}$ where $E_e$ is the energy of the observed electron which is determined based on the position in the Cherenkov detector array. Fig.~\ref{fig:bremrec} shows the number of electrons as function of the Cherenkov counter position. The Cherenkov detector array will consist of 15 detectors, each with a size of $2\times 2$~cm$^2$ spanning from 3.5~cm to 33.5~cm, covering the energies between 1 and 15~GeV. 
Also shown is the resolution of the photon energy reconstruction: a strong peak at $1$ is observed stemming from events with a single Bremsstrahlung photon. The tail to higher values comes from events where more than one photon has been radiated. The RMS of the distribution is 7\% after detector acceptance cuts ($E>10$~MeV and $|y|<5$~mm).

The positron energy spectrum is also related to the $\gamma_B$ energy as, on average, the positron energy is about $E_{\gamma_B}/2$. Using both measurements, the photon energy spectrum and the normalisation can be constrained. 

It is planned to test this technique in a test beam prior to the experiment. For such a testbeam, e.g.\ the prototype Cerenkov detector shown in Fig.~\ref{fig:cerdet} is already available and a stave of ALPIDE detectors and a prototype calorimeter could also be available. 

\subsubsection{Electron--positron Pair Production after the Interaction Point}
At the IP, electron--positron pairs will be produced both in $\gamma_B + n\gamma_L$ and in the one- and two step trident processes in $e^- + n\gamma_L$ interactions.
In addition, electrons are produced in the $e^- + n\gamma_L$ process. A spectrometer system is again 
employed, using a dipole magnet, capable of delivering a $B$-field up to 2.2~T, to separate the electrons and positrons from each other and from photons which continue down the beam line.  
After the magnet, a high-precision detector system is proposed in which several planes of silicon pixel detectors will be used for  track reconstruction.  
This will be complemented by calorimeters. The calorimeter will allow matching of tracks from the silicon detectors to energy deposits and hence provide extra information in the measurement of the electron and positron energy spectra. 

The electron and positron detectors in this case are expected to be identical for the $\gamma_B + n\gamma_L$ setup as the rates for both are expected to be the same. 

For the $e$--laser setup, the positron detector will stagger two staves on top of each other due to the larger spread in the vertical direction expected for the two-step trident process (see Fig.~\ref{fig:occ}). For electrons the rates are now very high and instead of a tracking system Cherenkov counters will be used, again followed by a calorimeter to measure the energy and to absorb the electrons. 

\begin{figure}[htp]
   \includegraphics[width=0.49\textwidth]{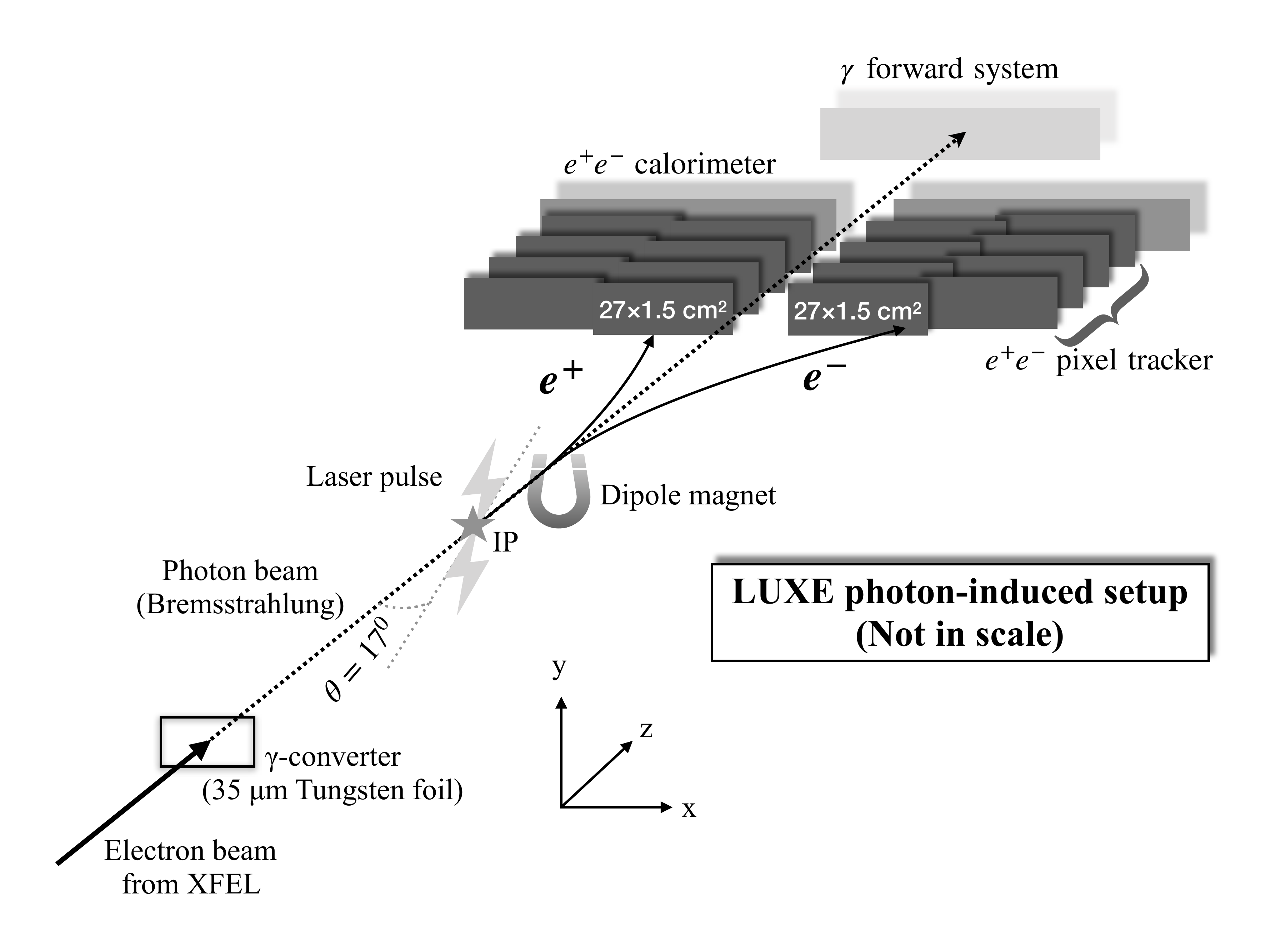}
   \includegraphics[width=0.49\textwidth]{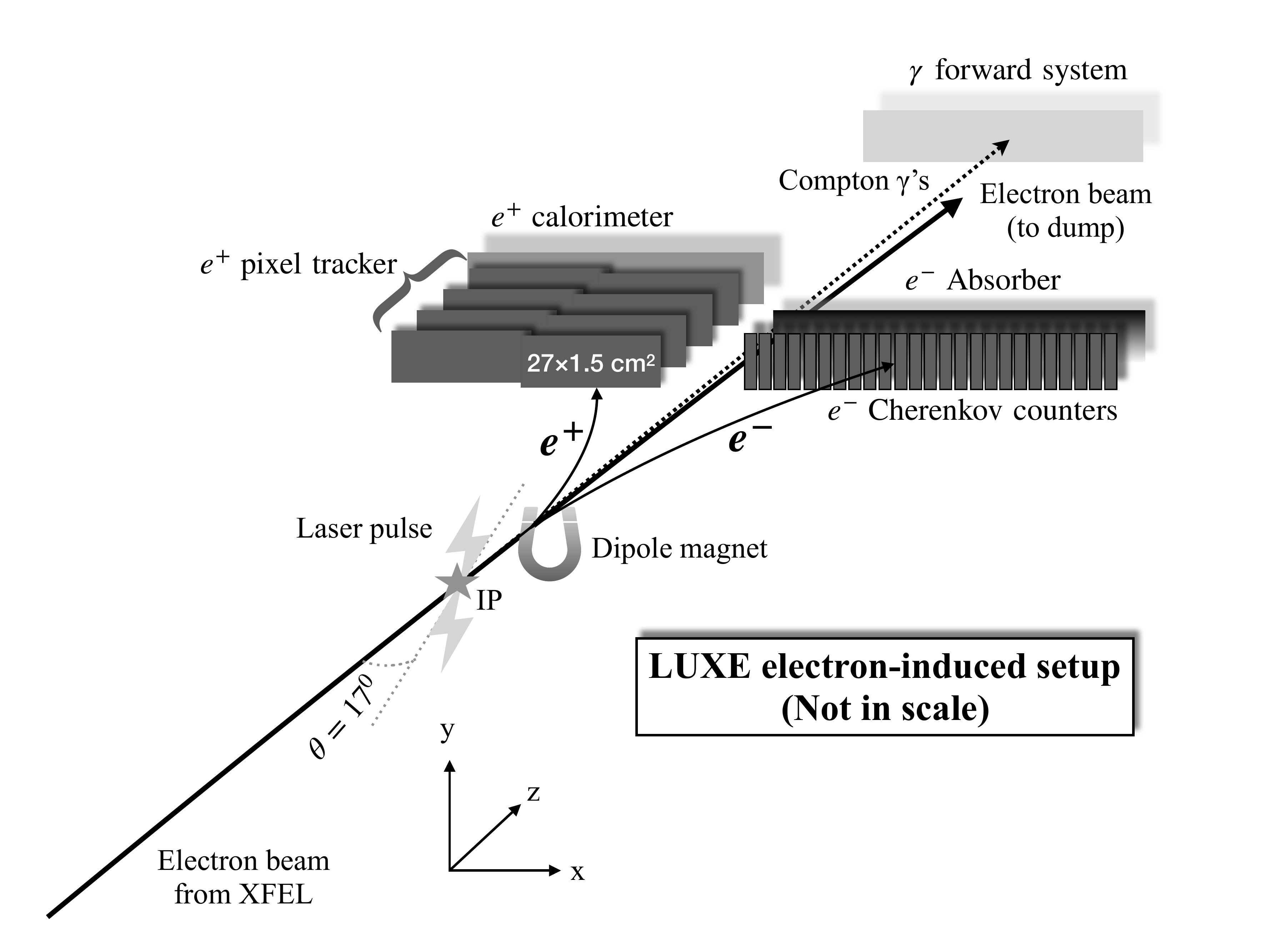}
   \caption{Schematic of the detector system to measure $e^+e^-$ pairs produced at the IP for the (left) $\gamma_B$--laser and (right) $e$--laser setup.  In both cases, the electrons and positrons are separated by a dipole magnet. In the $\gamma_B$--laser setup, electrons and positrons are measured in a detector system consisting of tracking planes and calorimeters, where the electron and positron 
   detector systems are identical.  In the $e$--laser setup, electrons and positrons are measured in a detector system consisting of tracking planes and a calorimeter for the positrons and Cherenkov counters for the electrons. Photons continue to the forward system, which is discussion in Section~\ref{sec:forward-photon}.
   \label{fig:eplus-eminus-spec}}
\end{figure}

\begin{figure}[htp]
\begin{center}
\includegraphics[width=0.48\textwidth]{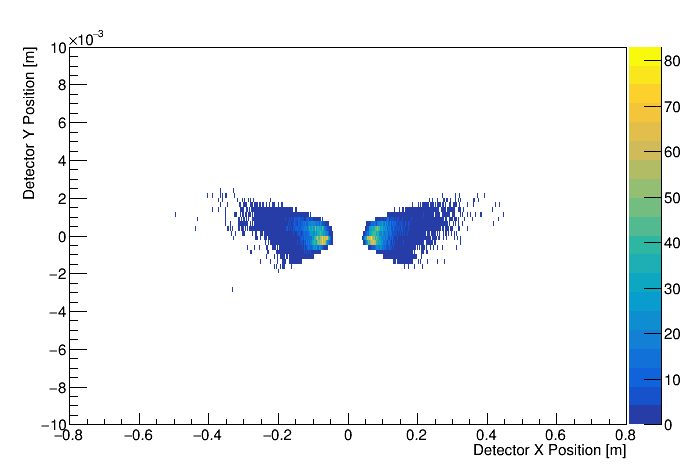}
   \includegraphics[width=0.48\textwidth]{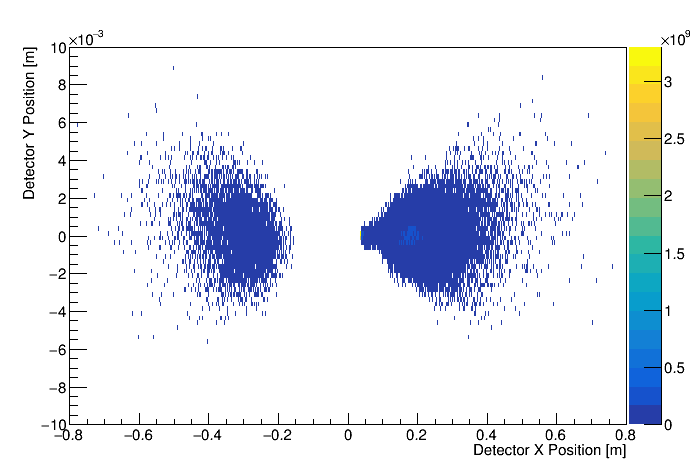}
   \caption{Occupancy of electrons and positrons as function of the transverse and vertical position for the process $\gamma_B+n\gamma_L \to e^+e^-$ after the dipole magnet with a field of 1.4~T, for $E_e=17.5$~GeV, nominal beam current and $\xi=2.5$. 
   The left figure shows the occupancy for the $\gamma_B$--laser setup and the right figure for the $e$--laser setup. Electrons are at positive $x$ and positrons at negative $x$. The normalisation of both figures is arbitrary. 
   \label{fig:occ}
   }
\end{center}
\end{figure}

A schematic of the setup is shown in Fig.~\ref{fig:eplus-eminus-spec}.

Fig.~\ref{fig:occ} shows the number of electrons and positrons as function of the impact position at the first silicon tracker plane for $\xi=2.5$ for the $\gamma_B$--laser and the $e$--laser setup. It is seen that they are in the horizontal plane in the range $5<|x|<60$~cm and in the vertical plane the spread is below $\pm 0.5$~cm for the $\gamma_B$--laser and $\pm 2$~cm for the $e$--laser setup. For the $\gamma_B$--laser setup. 

Based on this spread the detector geometry has been optimised. The horizontal size should be 50~cm to ensure that $e^-$ and $e^+$ can be measured at all energies in both setups. For the $e$--laser setup the vertical spread is larger as the intermediate photon may have a significant transverse momentum. Thus in this mode the vertical size of the detectors should be at least 3~cm while in the $\gamma_B$--setup 1~cm is sufficient.

\subsubsection{Forward Photon Spectrometer}
\label{sec:forward-photon}
\begin{figure}[htp]
   \includegraphics[trim={0cm 0cm 0cm 0cm},clip,width=\textwidth]{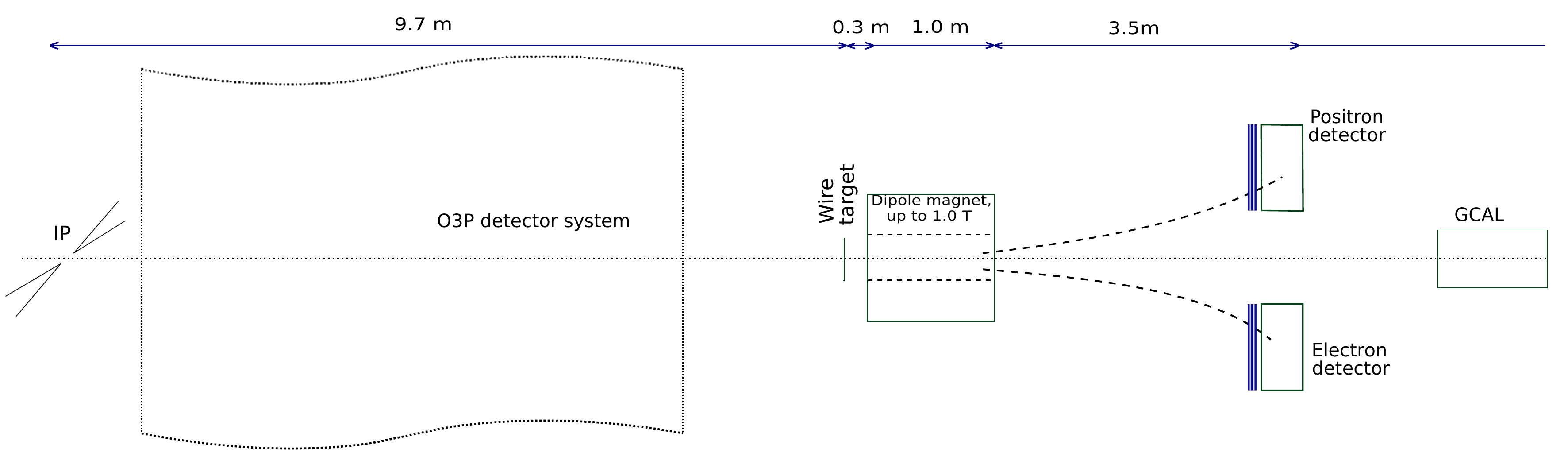}
   \caption{Schematic of the forward photon spectrometer.  Photons from the IP are incident on a wire target with some converting to $e^+e^-$ pairs which are then separated by a dipole magnet.  The spectra of electrons and positrons are measured in a detector system consisting of tracking planes and a calorimeter.  Photons that do not undergo conversion continue along the beam and are measured in a calorimeter. Note that for clarity instrumentation between the IP and wire target is indicated but not shown.}
   \label{fig:photon-spec-alt1}
\end{figure}

The final detector system is a forward photon spectrometer to detect photons from the high intensity Compton process, $e^- + n\gamma_L \to e^- \gamma$, 
but also to monitor photons which do not interact at the IP in the $\gamma_B + n\gamma_L$ process.  
The photons will pass through a wire target with a fraction converting to an electron--positron pair and the rest continuing along the beamline, where 
they will be measured in a calorimeter.   To measure the electron--positron pair, a magnetic spectrometer will again be employed and a detector consisting 
of tracking and a calorimeter, similar to the detector system just after the IP.  The use of a thin wire target ensures that a measurable number of $e^+e^-$ 
pairs are incident on the detector.  Simulations have been performed using a wire made of tungsten or nickel with a thickness of 10\,$\mu$m; and for $10^8$ initial photons produced about 150-4000 $e^+e^-$ pairs can be produced, depending on the wire thickness and material, in the range of energies $1 - 8$\,GeV which could be measured with precise detectors and effective particle reconstruction.  A schematic of the setup is shown in Fig.~\ref{fig:photon-spec-alt1}, including dimensions.

\subsection{Technology Solutions}
\subsubsection{Silicon Pixel Tracking Detectors}
\begin{figure}[htbp]
  \begin{minipage}[t]{0.48\textwidth}
    \includegraphics[width=0.95\textwidth]{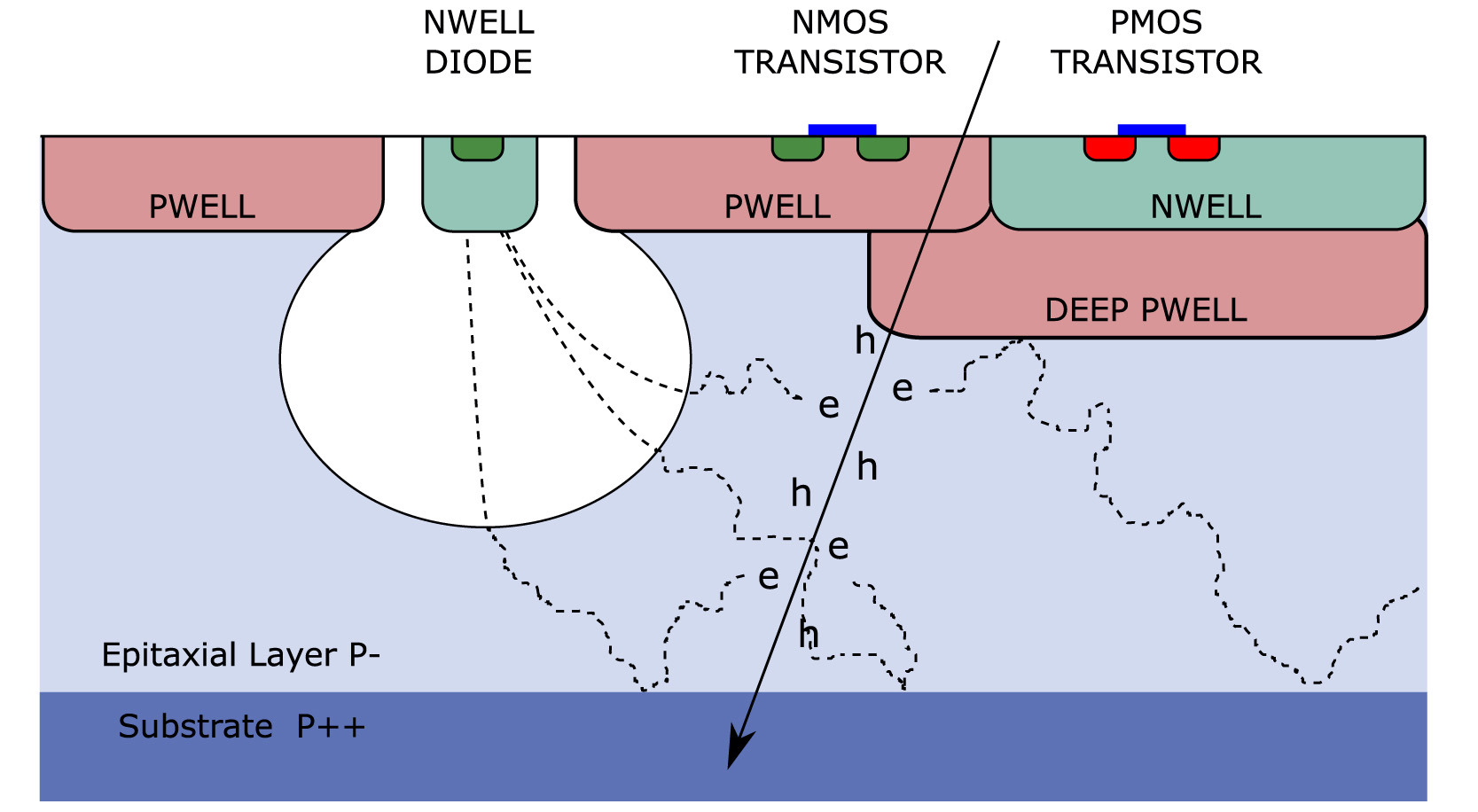}
    \caption{A schematic cross-section view of the ALPIDE pixel. The ionisation charge generated by the incident charged particle in the $25\,\mu$m thick epitaxial layer is collected by the n-well. A region in the epitaxial layer gets depleted as indicated in white by applying the bias voltage. For more details,  see~\cite{Senyukov:2013se,Mager:2016yvj}.}
    \label{fig_alpide_pixel}
  \end{minipage}\hfill
  \hspace{0.02\textwidth}
  \begin{minipage}[t]{0.48\textwidth}
    \includegraphics[trim={0cm 2cm 0cm 2cm},clip,width=0.99\columnwidth]{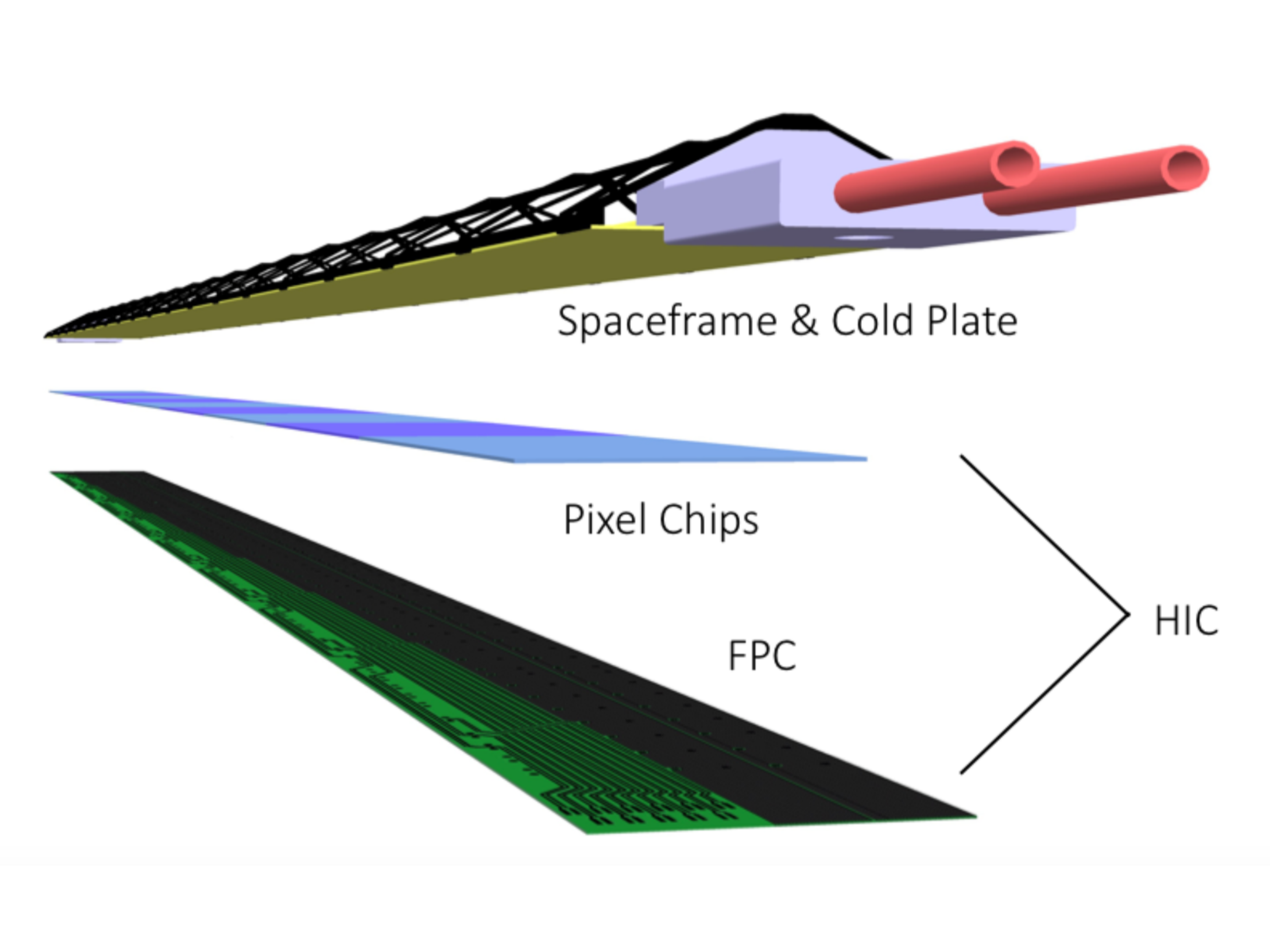}
    \caption{Schematic layout of a single stave. Nine pixel sensors are flip-chip mounted on a flexible printed circuit (FPC) to form a hybrid integrated circuit (HIC). The HIC is glued on a carbon fibre support structure (Spaceframe), which integrates a water cooling circuit (Cold Plate). For more details, see~\cite{ALICE-PUBLIC-2018-013}.}
    \label{fig_pixel}
  \end{minipage}\hfill
\end{figure}

The LUXE phase-0 tracking detector can be based almost entirely on the technology used for the upgrade of the inner tracking system of the ALICE experiment at the LHC~\cite{Abelevetal:2014dna,ALICE-PUBLIC-2018-013}.  
The LUXE tracker will be assembled out of four layers staggered at each of the two sides of the beamline downstream of the dipole magnet.  Each ``half-layer'' is built from nine $3 \times1.5$\,cm$^2$ ALPIDE monolithic active pixel sensors~\cite{Senyukov:2013se,Mager:2016yvj} lined up to cover an area of $27 \times 1.5$\,cm$^2$ with $100\,\mu$m gaps between adjacent sensors.  The ALPIDE sensors (see Fig.~\ref{fig_alpide_pixel}) pixel size is $27 \times 29\,\mu$m$^2$ resulting in a spatial resolution of $\sigma \sim 5$\,$\mu$m.  A time resolution of a few\,$\mu$s was demonstrated, with a very low probability for dark current appearance, orders of magnitude lower than for hybrid pixels.  These features are packed in a very low material budget of 0.05\%$X_0$ per layer (sensor only).

Each ``half-layer'' is built from a carbon fibre structure providing the mechanical support and the necessary stiffness (space frame), a sheet of high thermal-conductivity carbon fibre (in thermal contact with the pixel chips) with embedded polyimide cooling pipes (cold plate), which is integrated into the space frame, and a hybrid integrated circuit (HIC) consisting of a polyimide flexible printed circuit (FPC) onto which the pixel chips and some passive components are bonded. The interconnection between the pixel sensors and FPC is implemented via conventional aluminium wedge wire bonding.  The electrical substrate is the flexible printed circuit that distributes the supply and bias voltages as well as the data and control signals to the pixel sensors. The HIC is glued to the ``cold plate'' with the pixel chips facing it in order to maximise the cooling efficiency.  The polyimide cooling pipes embedded in the cold plate have an inner diameter of 1.024\,mm and a wall thickness of $25\,\mu$m.  These are filled with water during operation.  The overall structure mean thickness is $X/X_0 = 0.357$\%, with the material budget broken down to fractions of the total width as: $\sim 15$\% silicon sensor (with absolute thickness of $50\,\mu$m), $\sim 50$\% electrical substrate (FPC) including the passive components and the glue, $\sim 20$\% cooling circuit and $\sim 15$\% carbon spaceframe.  In LUXE, each ``layer'' comprises of two ``half-layers'' placed one behind another in the beam direction with a few centimetres of overlap in the transverse direction such that a total transverse length of $\sim 50$\,cm is achieved.  A cross section of a sensor and a schematic drawing of a ``half-layer'' is shown in Fig.~\ref{fig_pixel}.

Despite being a relatively new technology, ALPIDE-based assemblies are already planned in several near-future experiments, besides ALICE.  The performance, quality, availability, and low-cost of the end-point layers (including the sensors and all associated services) makes this solution particularly appealing for LUXE, with minimal adaptations.

\subsubsection{Calorimetry}
Calorimeters in combination with the spectrometers will provide LUXE with additional measurements to improve $e^+$, $e^-$ and Compton photon spectra 
reconstruction. In addition, calorimeters are an essential tool to reject low energy background generated in material in the experiment. These tasks require calorimeters with small transverse shower size, usually characterised by the Moli\`{e}re radius, and good energy resolution. 

\begin{figure}[h!]
  \begin{minipage}[t]{0.52\textwidth}
    \includegraphics[width=0.99\columnwidth]{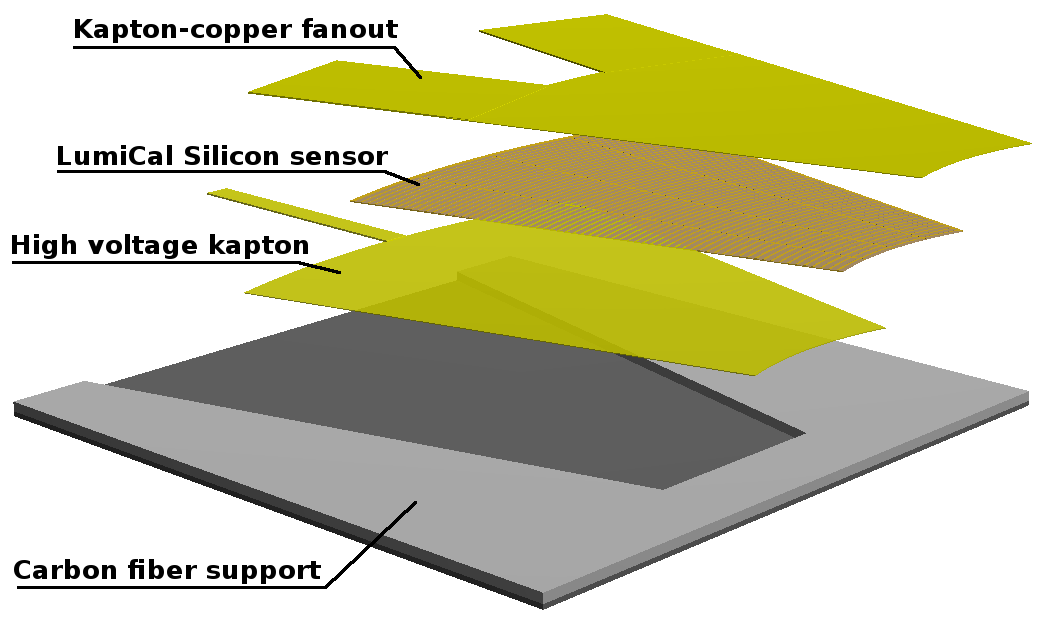}
    \caption{Detector plane assembly. The total thickness of the detector module is 650~$\mu$m.}
    \label{fig_ThinLCAssembly}
  \end{minipage}\hfill
  \hspace{0.025\textwidth}
  \begin{minipage}[t]{0.45\textwidth}
    \includegraphics[width=\textwidth]{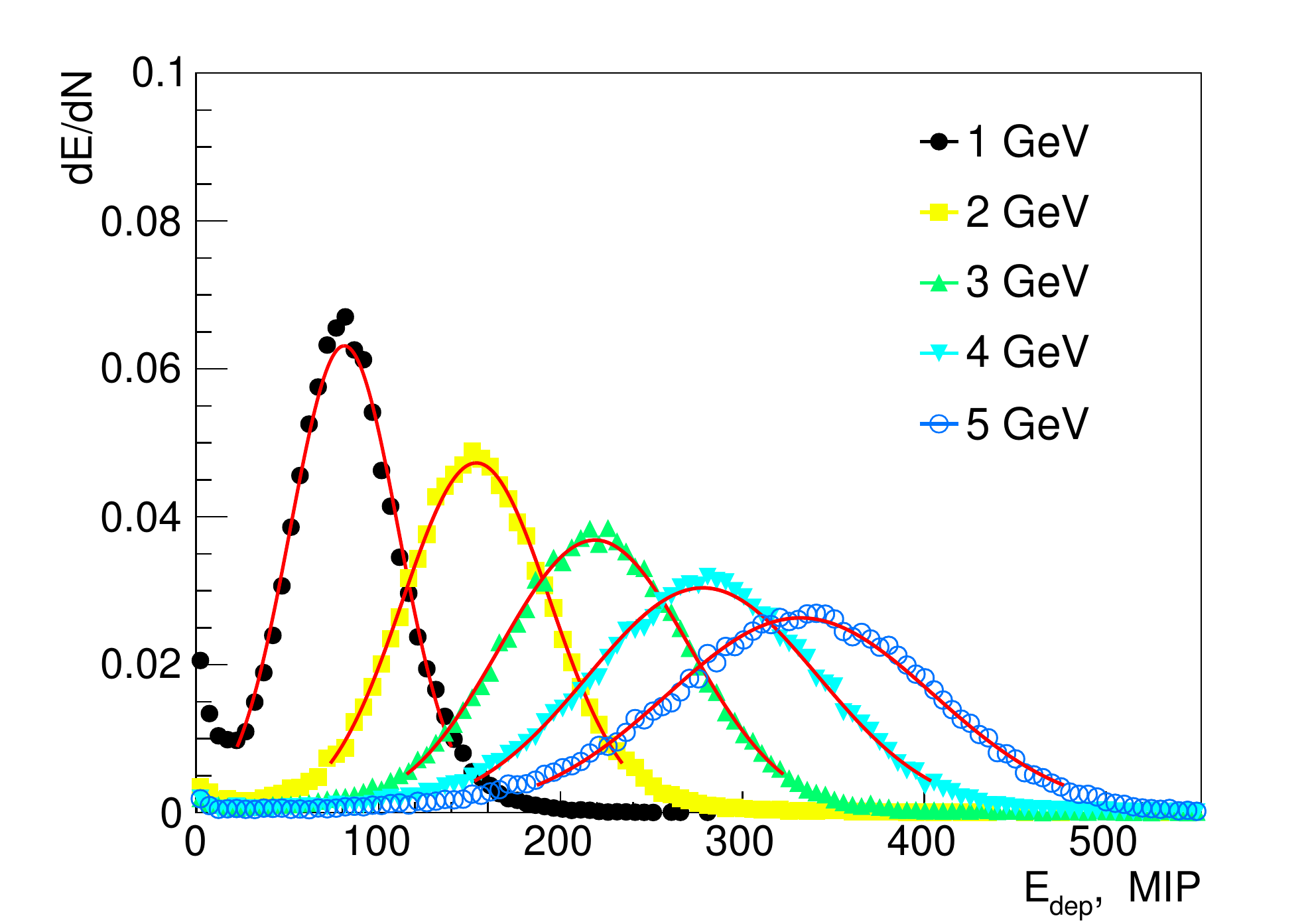}
    \caption{Distribution of the energy deposited in the calorimeter prototype, $dE/dN$, for different electron energies. The lines illustrate fits with a Gaussian.}
    \label{E_dep_total_1_5gev}
  \end{minipage}\hfill
\end{figure}

The ultracompact design of the LumiCal and BeamCal calorimeters foreseen in the very forward region of future International Linear Collider (ILC) and 
Compact Linear Collider (CLIC) experiments aims at solving similar problems. These include good efficiency for high energy electron and positron 
identification on top of a widely-spread background produced by beamstrahlung, high accuracy of position reconstruction and good energy resolution. 
The existing prototype of the LumiCal for future $e^+e^-$ colliders is a sampling calorimeter composed of 20\,layers of 3.5\,mm 
(1X$_0$) thick tungsten absorbers and silicon sensors placed in a 1\,mm gap between absorber plates. The pad sensor is made of a 320\,$\mu$m 
thick high resistivity n-type silicon wafer. A sketch of the structure of the detector plane is shown in Fig.~\ref{fig_ThinLCAssembly}. The bias voltage is supplied to the n-side of the sensor by a 70\,$\mu$m flexible Kapton--copper foil, glued to the sensor with a conductive glue. The 256\,pads of the sensor are connected to the front-end electronics using a fan-out made of 120\,$\mu$m thick flexible Kapton foil with copper traces. 
 Ultrasonic wire bonding is used to connect conductive traces on the fan-out to the sensor pads. A support structure, 
made of carbon fibre composite with a thickness of 100\,$\mu$m in the sensor-gluing area, provides mechanical stability for the prototype of the detector plane. 
The technology and assembly process were designed to ensure a sub-millimetre thickness of the module over the whole area of about 83\,cm$^2$ which allows its installation in the 1\,mm gap between tungsten absorber plates. 

The prototype of the compact calorimeter was tested at DESY with an electron beam in the range of $1 - 5$\,GeV and the resulting effective Moli\`{e}re radius at 5\,GeV was determined to be (8.1 $\pm$ 0.1 (stat) $\pm$ 0.3 (syst))\,mm, a value well reproduced by the MC simulation, (8.4 $\pm$ 0.1)\,mm. Further details can be 
found in~\cite{Abramowicz:2018vwb}. A small Moli\`{e}re radius of the calorimeter also means the decrease of its transverse size for a required fiducial volume.

Fig.~\ref{E_dep_total_1_5gev} shows the distributions of the energy deposited in the sensors of the calorimeter prototype measured in beam test
for electrons of different energies. Because of the limited number of layers installed in the prototype the energy leakage becomes notable as the energy of the particle increases.
Good energy resolution will be provided by the analog readout of the deposited energy with electronics whose dynamic range is extended to measure accurately both the central part of the electromagnetic shower with high energy deposition and its periphery where the signal can be at MIP (minimum ionising particle) level or smaller. This type of electronics has been designed and produced prototypes are planned to be tested with electron beam in the end of 2019.

\subsubsection{Cherenkov Detectors}
In any strong-field QED experiment, it will be important to monitor the quality of the interaction region and in particular the ability of the incoming beam electrons to actually traverse the strong-field region created by the focused high-power laser beam and indeed maintain stable strong-field interaction conditions 
throughout a typical data collection period.  An effective way of achieving this is through the use of Cherenkov monitors located downstream the IP dedicated to the detection of electrons within a narrow energy range indicative of the order of the non-linear Compton scattering which produced them. These monitors will also be used during the initial stages of the experiment to provide the necessary diagnostics in order to optimise the spatial and temporal overlap of the electron and laser beams.

The Cherenkov detectors can follow the design developed for Compton polarimetry at future lepton colliders, of which a 2-channel prototype has been built and successfully operated in testbeam~\cite{Bartels:2010eb}, and a calibration system for this detector has been described in~\cite{Vormwald:2015hla}. With the prototype and its calibration system, a linearity up to a few permille has been demonstrated over a dynamic range spanning a factor of $10^3$. As active medium a gas with a refractive index of only about one permille above unity (e.g.\ $C_4F_{10}$ with $n=1.0014$) has the advantage of being very robust against backgrounds from low-energy charged particles due to a Cherenkov threshold in the order of $10$\,MeV. If required by radiation conditions, the gas can be exchanged frequently or even cycled continuously. In principle also quartz can be considered as active medium~\cite{List:2015lsa}. With its about 200 times higher light yield per incoming electron, a quartz-based detector could be attractive for the locations in LUXE with lower rates, e.g.\ after the IP in the $\gamma_B$--laser setup. 

\begin{figure}[htbp] 
\begin{subfigure}{0.45\hsize} 
\includegraphics[width=\textwidth]{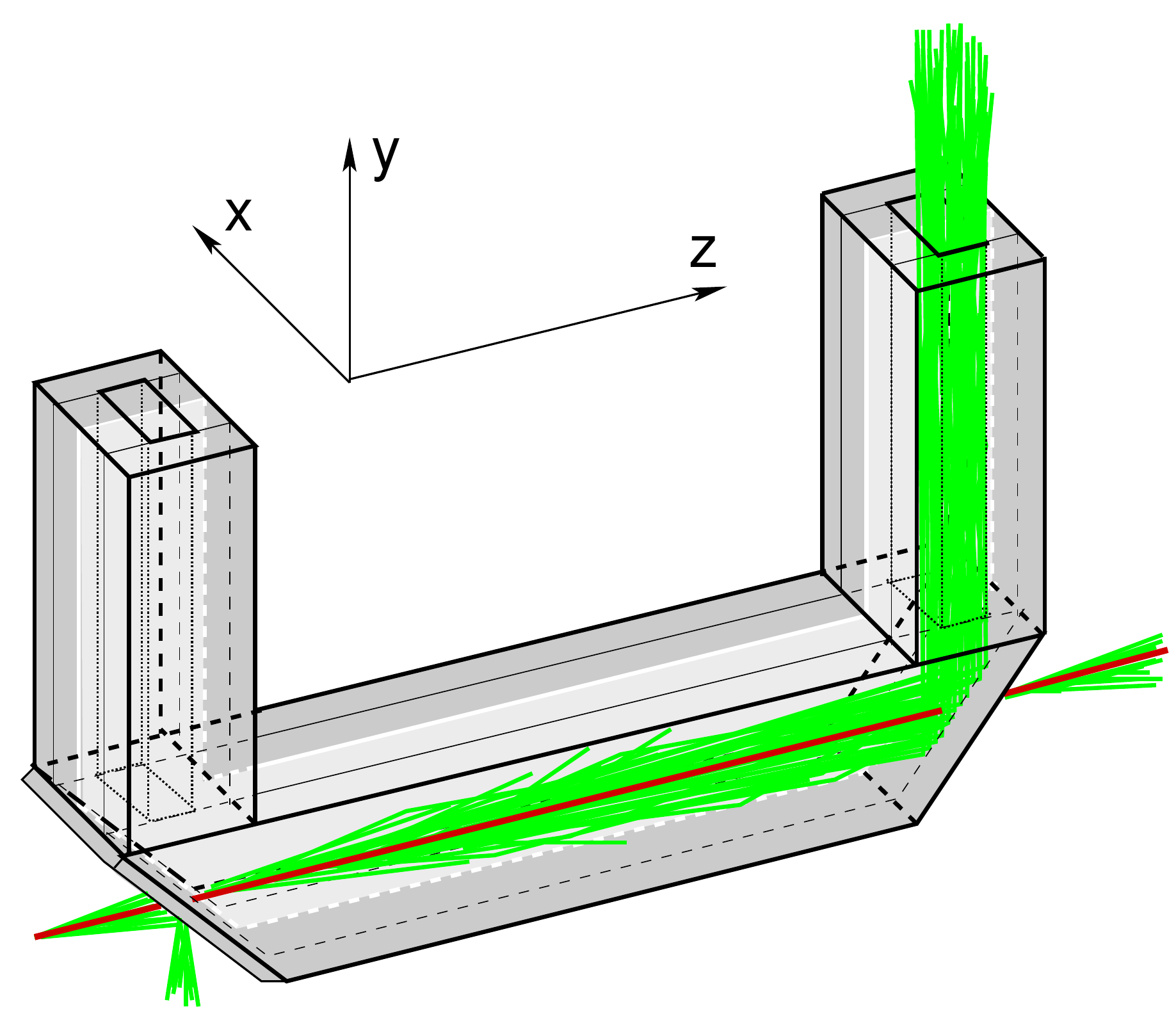}
\caption{ \label{fig:cerdet:simulation} Geant4 simulation of Cherenkov light created by an electron passing through the detector.}
\end{subfigure}
\hspace{0.6cm}
\begin{subfigure}{0.48\hsize} 
\includegraphics[width=\textwidth]{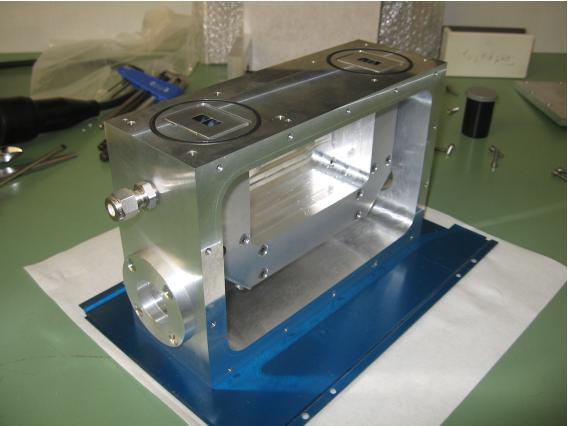}
\caption{  \label{fig:cerdet:assembly} Photo taken during the assembly of the actual prototype.}
\end{subfigure}
\caption{Two-channel prototype of a gas Cherenkov detector developed for Compton polarimetry~\cite{Bartels:2010eb}, which serves as basis for the Cherenkov detectors for LUXE.}
\label{fig:cerdet}
\end{figure}

Fig.~\ref{fig:cerdet:simulation} shows the two-channel prototype in a Geant4 simulation with an electron passing through the horizontal part of the gas-filled ``U''-shaped structure. A mirror in the corner reflects the Cherenkov light upwards to the photodetectors, which are placed outside of the plane of the main beam and the spectrometer dipole in order to minimise their radiation load. A photo taken during the assembly of the actual prototype is shown in Fig.~\ref{fig:cerdet:assembly}. The channels have a square cross section of about $1\times1$\,cm$^2$. This cross section can be increased by the factor of about two to match the granularity required for LUXE, and to the surface of the selected photodetectors. While in principle a single-anode PD would be sufficient, it has been shown in the testbeam campaign that photodetectors with a segmented anode can be very useful in order to align the detector~\cite{Bartels:2010eb}. 

The LUXE requirements on linearity and homogeneity of the response as well as on the precision of the alignment are somewhat less challenging than for polarimetry, while the requirements on the dynamic range, on background tolerance as well as on radiation hardness will be similar or even more stringent. Nevertheless, the similarity is sufficiently large to be confident that the existing design can be adjusted with reasonable effort.

\subsection{Data Acquisition and Control System}
As the maximum data-taking frequency will be  a 10~Hz and all detectors are small, current data acquisition (DAQ) solutions should be appropriate for LUXE.  Calibration 
data will also be needed, e.g.\ when there is an electron bunch but no electron--laser events or when there are no electron bunches, in order to measure 
pedestals, noise and backgrounds.  This again should not yield huge data rates or high data volume.  The DAQ system will need to be bi-directional as control data 
will need to be sent to the detectors, e.g. to distribute timing information, to control motorised detector stages, etc..

Therefore each detector component, e.g.\ silicon tracker after the IP or forward photon calorimeter (GCAL) can be read out and controlled by one front-end computer 
(PC) which is interfaced to the front-end electronics.  Note more than one PC may be used for contingency, although the actual data rates would not require this.  
Each of the detector PCs will then be connected to a central DAQ PC as well as a trigger and control system.  Further PCs will be required to display detector 
information for e.g.\ data quality monitoring purposes.

A DAQ software will be required which could be used for individual components or if they come with their own software a central DAQ software will be needed to interface to them.  Many different DAQ softwares exist with different levels of complexity and scale, with some designed for specific experiments and others as generic developments. An example software that could be used is EUDAQ2~\footnote{\tt https://eudaq.github.io} which has been developed for tests of high energy physics prototype detectors, which provide a similar setup to LUXE. It can cope with different triggering and readout schemes and has been used by several different detectors and projects.

\subsection{Installation and Commissioning}
The installation and commissioning need to be prepared very well as access to the tunnel will be limited. It is planned to use robotics to ensure they can be moved and aligned within the tunnel after installation. The three independent detector systems (after tungsten foil, after IP and photon detection) will be built up prior to installation independently at DESY, and can be commissioned largely on the surface before installing them in the tunnel. They can also be tested in test beam (in parts and in their entirety) to calibrate and commission them. 

\section{Schedule and Milestones}
\textbf{Winter 2019/2020 and 2020/2021:} Installation is assumed to extend over two winter shutdowns \\
\textbf{2022:} prototype experiment with 30 TW laser in $e$--laser setup. 
Commissioning, data taking and publication of results\\
\textbf{2023:} prototype experiment with 30 TW laser in $\gamma_B$--laser setup. 
Commissioning, data taking and publication of results\\
\textbf{2024:} Install 300~TW laser \\
\textbf{2025-2027:} Commissioning an data taking with 300~TW laser in $e$--laser and $\gamma_B$--laser in subsequent years \newline

During data taking, it is planned to benefit from the beam configurations used for photon science. It is interesting to explore different electron energies and ideally data will be taken for about one month at a given energy. 

\section{Conclusions and Outlook}
In this Letter we have presented the science case and the technical aspects of a newly proposed experiment to study interactions of high-energy electrons and photons in a strong laser field. The experiment proposed plans to use the electron beam of the European XFEL and is called LUXE, and has the potential to pioneer an new regime of quantum physics. The technical aspects of the experiment are challenging but solutions exist already in other particle physics, photon science or laser experiments, and can be applied. Based on initial simulations the LUXE experiment will be able to make the measurements envisaged with the proposed setup. 

\section*{Acknowledgements}
We thank the members of various DESY groups without whom this work would not have been possible: MVS (vacuum modification), MIN (kicker magnet, beam dump), D3 (radio protection), MEA (installation and magnets), ZM1 (construction), MKK (power and water), IPP (CAD integration).

We also thank the DESY directorate for funding this work through the DESY Strategy Fund. The work by Dr. Beate Heinemann was in part funded by the Deutsche Forschungsgemeinschaft under Germany‘s Excellence Strategy – EXC 2121 ``Quantum Universe" – 390833306. The work by Dr. A. Hartin and Dr. M. Wing was supported by the Leverhulme Trust Research Project Grant RPG-2017-143 in the UK. The work of Dr. Gilad Perez is supported by grants from the BSF, ERC-COG, ISF, Minerva, and the Segre Research Award. Dr. Gianluca Sarri wishes to acknowledge support from EPSRC (grant Nos: EP/N027175/1 and EP/P010059/1).

This work has benefited from computing services provided by the German National Analysis Facility (NAF).

\bibliography{Biblio}

\end{document}